\documentclass[aip,jcp,superscriptaddress,showpacs,floatfix,preprint]{revtex4-1} 
\usepackage{amsmath}
\usepackage{amssymb}
\usepackage{graphicx}
\usepackage{makeidx}

\newcommand{\bra}[1]{\left \langle #1 \right \vert}
\newcommand{\ket}[1]{\left \vert #1 \right \rangle}

\begin{document}
\title{A Unified Stochastic Formulation of Dissipative Quantum Dynamics. I. Generalized Hierarchical Equations}

\author{Chang-Yu Hsieh}
\affiliation{Department of Chemistry, Massachusetts Institute of Technology,
77 Massachusetts Avenue, Cambridge, MA 02139}
\affiliation{Singapore University of Technology and Design, Engineering Product Development, 8 Somapah Road, Singapore 487372}
\author{Jianshu Cao}
\affiliation{Department of Chemistry, Massachusetts Institute of Technology,
77 Massachusetts Avenue, Cambridge, MA 02139}
\affiliation{Singapore-MIT Alliance for Research and Technology (SMART) Centre, Singapore 138602}

\begin{abstract}
We extend a standard stochastic theory to study open quantum systems coupled to generic quantum environments
including the three fundamental classes of  non-interacting particles: bosons, fermions and spins.
In this unified stochastic approach, the generalized stochastic Liouville equation (SLE) formally captures the exact 
quantum dissipations when noise variables with appropriate statistics for different bath models are applied. 
Anharmonic effects of a non-Gaussian bath are precisely encoded in the bath multi-time correlation functions that noise variables have
to satisfy.  Staring from the SLE, we devise a family of generalized hierarchical equations by averaging out the noise variables
and expand bath multi-time correlation functions in a complete basis of orthonormal functions.
The general hiearchical equations constitute systems of linear equations that provide numerically exact simulations
of quantum dynamics.  For bosonic bath models, our general hierarchical equation of motion reduces exactly to an extended version of hierarchical equation of motion which allows efficient simulation for arbitrary spectral densities and temperature regimes.
Similar efficiency and flexibility can be achieved for the fermionic bath models within our formalism.
The spin bath models can be simulated with two complementary approaches in the presetn formalism.
(I) They can be viewed as an example of non-Gaussian bath models and be directly handled 
with the general hierarchical equation approach given their multi-time correlation functions.  
(II) Alterantively, each bath spin can be first mapped onto a pair of fermions and be treated as fermionic environments within the present formalism.   
\end{abstract}

\maketitle

\section{Introduction}\label{sec:intr}

Understanding dissipative quantum dynamics of a system embedded in a complex environment is an important 
topic across various sub-disciplines of physics and chemistry.  Significant progress in the understanding of 
condensed phase dynamics have been achieved within the context of a few prototypical models\cite{Leggett:1987wk,breuer:book,weiss:book} 
such as
Caldeira-Leggett model and spin-boson model.  In most cases
the environment is modeled as a bosonic bath, a set of non-interacting harmonic oscillators 
whose influences on the system is concisely encoded in a spectral density.
The prevalent adoption of bosonic bath models is based on the arguments 
that knowing the linear response of an environment near 
equilibrium should be sufficient to predict the dissipative quantum dynamics of the system.

Despite many important advancements in the quantum dissipation theory
have been made with the standard bosonic bath models in the past decades, 
more and more physical and chemical studies have suggested the essential roles
that other bath models assume.  We briefly summarize three scenarios below.

\begin{enumerate}
\item{ A standard bosonic bath model fails to predict the correct electron transfer rate in donor-acceptor complex strongly coupled to some low-frequency intramolecular modes.
Some past attempts to model such an anharmonic, condensed phase environment include (a) using a bath of non-interacting Morse\cite{Wang:2007eu,LopezLopez:2011en,Kryvohuz:2005jf,wu_cao_jcp01} or quartic oscillatorsand (b) mapping anharmonic environment onto effective harmonic modes\cite{cao_voth_jcp95,makri99} with a temperature-dependent spectral density.} 

\item{
Another prominent example is the fermonic bath model. Electronic transports through nanostructures, such as quantum dots or molecular junctions, involves particle exchange occurs across the system-bath boundary.
Recent developments of several many-body physics and chemistry methods, such as the dynamical mean-field theory\cite{dmft06} and the density matrix embedding theory\cite{dmet12}, 
reformulate the original problem in such a way that a crucial part of the methods is to solve an open quantum impurity model embedded in a fermionic environment.}

\item{The spin (two-level system) 
bath models have also received increased attention over the years due to ongoing interests in developing various solid-state quantum technologies\cite{hsieh_2012} under the ultralow temperature
when the phonon or vibrational modes are frozen and coupling to other physical spins (such as nuclear spins carried by the lattice atoms), impurities or defects in the host material 
emerge as the dominant channels of decoherence.} 
\end{enumerate}     

Both bosonic and fermionic environments are Gaussian baths, which can be exactly treated by the linear response\cite{makri99} in the path integral formalism.  
For the non-Gaussian baths, attaining numerically exact open quantum 
dynamics would require either access to higher order response function of the bath in terms of its multi-time correlation functions or explicit dynamical treatments of the bath degrees of freedom (DOFs).  

In this work, we extend a stochastic formulation\cite{Stockburger:2002em,Lacroix:2005in,Shao:2004et} of quantum
dissipation by incorporating all three fundamental bath models: non-interacting 
bosons, fermions and spins.  
The stochastic Liouville equation (SLE), Eq.~(\ref{eq:sleq}), 
prescribes a simple yet general form of quantum dissipative dynamics when the bath effects are modelled as colored noises $\xi(t)$ and $\eta(t)$.
Different bath models and bath properties are distinguished in the present framework by assigning distinct noise variables and associated statistics. 
For instance, in dealing with bosonic and fermionic baths, the noises are complex-valued and Grassmann-valued Gaussian processes, respectively, and characterized by the two-time correlation functions such as
Eq.~(\ref{eq:xi_corr}).  
The Grassmann-valued noises are adopted whenever the environment is composed of fermionic modes as these algebraic entities would bring out the Gaussian characteristics of fermionic modes.  
For anharmonic environments, such as a spin bath, the required noises are generally non-Gaussian.
Two-time statistics cannot fully distinguish these processes and higher order statistics furnished with bath
multi-time correlation functions are needed.

Despite the conceptual simplicity of the SLE, achieving stable convergences in stochastic simulations
has proven to be challenging in the long-time limit. 
Even for the most well-studied bosonic bath models, it is still an active research topic to develop
efficient stochastic simulation schems\cite{stockburger_epl2016,wiedmann_stockburger_pra2016,zhou_chen_pra2016} today.
Our group has successfully applied stochastic path integral simulations
to calculate (imaginary-time) thermal distributions\cite{moix_prb2012}, 
absorption/emission spectra\cite{Moix:2015ei} and energy transfer\cite{Moix:2015ei,Moix:2013jb};
however, a direct stochastic simulation of real-time dynamics remains formidable. 
In this study, we consider generic quantum environments that either exhibit non-Gaussian characteristics or
invovles ferimonic degrees of freedoms (and associated Grassmann noise in the stochastic formalism).  
Both scenarios present new challenges to developing efficient stochastic simulations. 
Hence, in subsequent discussions, all numerical methods developed are strictly deterministic.  We note that it is common to  derive exact master equation\cite{Chen_You_pra2013,Li:2011dj}, hierarchical equation of motions\cite{vega2015,Zhou:2007fx}, and hybrid stochastic-deterministic numerical methods\cite{moix_prb2012,Moix:2013jb,Zhou:2007fx} from a stochastic formulation of open quantum theory. 
In sec.~\ref{sec:spectral}, we further illustrate the usefulness of our stochastic frmulation by presenting
a numerical scheme that would be difficult to obtain within a strictly deterministic framework of open quantum theory. Furthermore, the stochastic formalism gives straightforward prescriptions to compute dynamical 
quantities such as $\left \langle F_s Q_B \right\rangle$, which represents system-bath joint observables, 
as done in a recently proposed theory\cite{Yan:2014cc}. 

Staring from the SLE, we derive numerical schemes to deterministically simulate quantum 
dynamics for all three fundamental bath models. 
The key step is to formally average out the noise variables in the SLE.
A common approach is to introduce auxiliary density matrices (ADMs), in close parallel to
the hierarchical equation of motion (HEOM) formalism\cite{Tanimura:2006ga,*ishizaki_tanimura_jpsp2005}, 
that fold noise-induced fluctuations on reduced density matrix
in these auxiliary constructs with their own equations of motions.  
To facilitate the formal derivation with the noise averaging,
we consider two distinct ways to expand the ADMs with respect to a complete set of orthonormal functions in the time 
domain. In the first case, the basis set corresponds to the eigenfunctions of the bath's two-time correlation functions. This approach provides an efficient description of open quantum dynamics for bosonic bath models.
Unfortunately, it is not convenient to extend this approach to study non-Gaussian bath models.  
We then investigate another approach inspired by a recent work on the extended HEOM (eHEOM)\cite{Tang:2016gh}.  
In this case, we expand the bath's multi-time correlation functions in an arbitrary set of orthonormal functions.
This approach generalizes eHEOM to the study of non-Gaussian and fermionic bath models with arbitrary spectral 
densities and temperature regimes.  Despite having a slightly more complex form, our fermionic HEOM can be easily related to the existing formalism\cite{yyj_wiley16}.
In this work, we refer to the family of numerical schemes discussed in this work collectively as the generalized 
hierarchical equations (GHE).

Among the three fundamental bath models, spin baths deserve more attentions.
A spin bath can feature very different physical properties\cite{Prokofiev:420342} from 
the standard heat bath composed of non-interacting
bosons; especially, when the bath is composed of localized nuclear / electron spins\cite{hsieh_2012,Kloeffel:2013eg}, defects and impurities.  This kind of spin environment is often of a finite-size and has 
an extremely narrow bandwidth of frequencies.
To more efficiently handle this situation, we consider a dual-fermion mapping that transforms each
spin into a pair of coupled fermions.  At the expense of introducing an extra set of fermionic DOFs, it becomes
possible to recast the non-Gaussian properties of the original spin bath in terms of Gaussian processes in the 
extended space. In a subsequent work, the paper II\cite{hsieh_cao_jcp16}, we should further investigate physical
properties of spin bath models.  

The paper is organized as follows. In Sec.~\ref{sec:2}, we introduce thoroughly 
the stochastic formalism for open quantum systems embedeed in a generic quantum environment.
The SLE is the starting point that we build upon to construct generalized hierarchical equations (GHE), a family
of deterministic simulation methods after formally averaging out the noise variables.
In Sec.~\ref{sec:spectral}, we study bosonic baths by expanding the noise processes in terms of the spectral eigenfunctions of the bath's two-time correlation function.  
In Sec.~\ref{sec:heom}, we discuss the alternative derivation that generalizes the recently introduced 
extended HEOM to study non-Gaussian bath models and fermionic bath models.
In Sec.~\ref{sec:spin}, we introduce the dual-fermion representation and 
derive an alternative GHE more suitable for spin bath models composed of nuclear / electron spins.
A brief summary is given in Sec.~\ref{sec:last}. In App.\ref{app:stochprocess}-\ref{app:grassmann}, we provide additional materials on the stochastic calculus and Grassmann number to clarify some details of the present work.  
Appendix \ref{app:inf-func} shows how to recover the influence functional theory from the present stochastic formalism.  In the last appendix, we present numerical examples to illustrate the methods discussed 
in the main text.

\section{A Unified Stochastic Framework For Open Quantum Systems}\label{sec:2}
\subsection{The system and bath model}
In this study, we consider the following joint system and bath Hamiltonian,
\begin{eqnarray}\label{eq:genH}
\hat H  & = & \hat H_s + \hat H_B + \hat H_{\text{int}} \nonumber \\
 & = & \hat H_s + \hat H_B + \sum_{\alpha=\pm} A^\alpha B^{\bar \alpha}, 
\end{eqnarray}
where the interacting Hamiltonian can be decomposed into factorized forms with $A^{\alpha}$ and $B^{\bar\alpha}$
acting on the system and bath, respectively.
More specifically, we assume $H_B = \sum_{k>0} \omega_k b^{\dag}_k b_k$ and
$B^{\alpha} =\sum_{k>0} g_k b^{\alpha}_{k}$ where  
the operator $b^{\dag}_{k}$ can be taken as the bosonic creation operator $a^{\dag}_k$, 
the fermionic creation operator $c^{\dag}_k$ or 
the spin raising operator $\sigma^+_k$ depending on the specific bath model considered. 
Throughout the rest of this paper, the system Hamiltonian reads
\begin{eqnarray}\label{eq:sysH}
\hat H_s = \frac{\omega_0}{2} \hat \sigma^z_0 + \frac{\Delta}{2} \hat \sigma^x_0,
\end{eqnarray}
where we simply take the system as a spin and the index 0 always refers to the system.
Although we adopt a specific system Hamiltonian in Eq.~(\ref{eq:sysH}), it can be more general.

We consider the factorized initial conditions,
\begin{eqnarray}\label{eq:genrhoi}
\hat \rho(0) & = & \hat \rho_s(0) \otimes \hat \rho_B^{\text{eq}},
\end{eqnarray}
where two parts are initially uncorrelated and 
the bath density matrix is commonly taken as a tensor product of thermal states for each individual mode.
With Eqs.(\ref{eq:genH})-(\ref{eq:genrhoi}), the dynamics of the composite system (system and bath) 
is obtained by solving the von Neumann equation
\begin{eqnarray}\label{eq:vneq}
\frac{\partial \hat \rho}{\partial t} = -i \left[ \hat{H}, \hat{\rho} \right].
\end{eqnarray}
If only the system is of interest, then one can trace over the bath DOFs,
i.e. $\hat\rho_s(t) = \text{Tr}_B \hat\rho(t)$.  
This straightforward computation (full many-body dynamics then partial trace) soon becomes 
intractable as the dimension of the Hilbert space scales exponentially with respect to a possibly large number of bath modes.

\subsection{Stochastic Decoupling of Many-Body Quantum Dynamics}

Many open quantum system techniques have been proposed to avoid a direct computation of Eq.~(\ref{eq:vneq}). 
Our starting point is to replace Eq.~(\ref{eq:vneq}) with a set of coupled stochastic differential equations,
\begin{eqnarray}
\label{eq:ito_unnorm}
& & d\tilde\rho_s  = \\
& & -i dt \left[\hat{H}_s, \tilde\rho_s\right] 
-\frac{i}{\sqrt 2} \sum_{\alpha=\pm} A^\alpha\tilde\rho_s dW_\alpha 
+ \frac{i}{\sqrt 2}\sum_{\alpha=\pm} \tilde\rho_s A^\alpha dV_\alpha, \nonumber \\
\label{eq:ito_unnorm2}
& & d\tilde\rho_B   =  \\
& & -i dt \left[\hat{H}_B, \tilde\rho_B\right] 
+\frac{1}{\sqrt 2} \sum_{\alpha=\pm} dW^*_\alpha B^{\bar \alpha}  \tilde \rho_B  
+ \frac{1}{\sqrt 2}\sum_{\alpha=\pm} dV^*_\alpha \tilde \rho_B  B^{\bar \alpha}, \nonumber
\end{eqnarray}
where the stochastic noises appear as differential Wiener increments, $dW_\alpha(t) = \mu_\alpha(t) dt $ and $dV_\alpha = \nu_\alpha(t)dt$ (the white noises $\mu_\alpha(t)$ and $\nu_\alpha(t)$ will be explicitly defined later).
Noting the stochastic relation\cite{cao_ungar_jcp96,Lacroix:2005in,Diosi:1998px} between density matrix and wave function, $\tilde\rho_s(t) = \ket{\psi^+(t)}\bra{\psi^-(t)}$, the two forward/backward wave functions
evolves under the time-dependent Hamiltonian, $H^{+}(t) = H_s + A\mu(t)$ and $H^-(t)= H_s + A^\dag \nu^*(t)$, respectively.  This connection implies an equivalent formulation of stochastic wave function-based theory of open quantum systems with better scaling\cite{zhong_cao_14} for large system simulations. In this work, we focus on the density matrix presentation and attach the tilda symbol on top of all stochastically-evolved density matrices 
as in Eqs.~(\ref{eq:ito_unnorm})-(\ref{eq:ito_unnorm2}).

Equation (\ref{eq:ito_unnorm2}) can be further decomposed into corresponding equations for individual modes,
\begin{eqnarray}\label{eq:ito_unnorm3}
& & d\tilde \rho_k = -i dt \left[H_{b,k} , \tilde\rho_k\right]   \nonumber \\
& & 
+\frac{1}{\sqrt 2} \sum_{\alpha=\pm} g_k dW^*_\alpha  b^{\bar\alpha}_k \tilde \rho_k  
+ \frac{1}{\sqrt 2}\sum_{\alpha=\pm} g_k dV^*_\alpha \tilde \rho_k b^{\bar\alpha}_k,
\end{eqnarray}
where $k=1, \dots, N_B$, $H_{b,k}=\omega_k b^{\dag}_k b_k$
 and $\tilde \rho_B = \otimes_{k=1}^{N_B} \tilde\rho_k$.
The system and bath are decoupled from each other 
but subjected to the same set of random fields.  
The stochastic processes ($W_\alpha(t)$ and $V_\alpha(t)$) in this work can be either complex-valued 
or Grassmann-valued depending on the bath model under study. 
To manifest the Gaussian properties of fermionic baths, 
it is essential to adopt the Grassmann-valued noises.  In these cases, 
it is crucial to maintain the order between fermionic operators and Grassmann-valued noise variables presented in Eqs.~(\ref{eq:ito_unnorm2})-(\ref{eq:ito_unnorm3}) as negative signs arise when the order of 
variables and operators are switched.

The white noises satisfy the standard relations
\begin{eqnarray}\label{eq:noise}
& & \overline{ \mu_\alpha(t)} = \overline{\nu_\alpha(t)} = 0, \nonumber \\
& & \overline{ \mu_\alpha(t) \mu^*_{\alpha'} (t^\prime)} = \overline{ \nu_\alpha(t) \nu^*_{\alpha'} (t^\prime)}  = 2\delta_{\alpha,\alpha'}\delta(t-t^\prime),
\end{eqnarray}
where the overlines denote averages over noise realizations.  
Any other unspecified two-time correlation functions vanish exactly. 
The order of variables in Eq.~(\ref{eq:noise}) also matters for Grassmann noises as explained earlier. 
With these basic properties laid out, we elucidate how to recover 
Eq.~(\ref{eq:vneq}) from the stochastic formalism.  
First, the equation of motion for the joint density matrix $\tilde\rho(t) = \tilde \rho_s(t) \tilde \rho_B(t)$ is 
given by
\begin{eqnarray}\label{eq:itodiff}
d\tilde \rho(t) = [ d\tilde \rho_s(t) ] \tilde \rho_B(t) + \tilde \rho_s(t) [ d\tilde \rho_B(t)]
+ d \tilde \rho_s(t) d \tilde \rho_B(t),
\end{eqnarray}
where the last term is needed to account for all differentials up to $\mathcal{O}(dt)$ as 
the product of the conjugate pairs of differential Wiener increments such as $\overline{dW_\alpha(t) dW^*_\alpha(t)} = 2 dt$,  contributes a term 
proportional to $dt$ on average.
Taking the noise averages of Eq.~(\ref{eq:itodiff}),
the first two terms together yield $-i dt [\hat{H}_s + \hat{H}_B, \overline{\tilde\rho_s \tilde\rho_B} ]$ and the last term gives the system-bath 
interaction, $-idt[\hat{H}_{I},\overline{\bar \rho_s \bar \rho_B}]$.  Due to the linearity of the von Neumann equation and the factorized initial 
condition, the composite system dynamics is given by $\hat \rho(t) = \overline{\tilde \rho_s(t) \tilde \rho_B(t)}$.  
To extract the reduced density matrix, we trace out the bath DOFs before taking the noise average,
\begin{eqnarray}\label{eq:itotrace}
\hat \rho_s(t) = \overline{\tilde \rho_s(t) \text{Tr}_B \tilde \rho_B(t)}.
\end{eqnarray}
In this formulation, it is clear that all the bath-induced dissipative effects are encoded in 
the trace of the bath's density matrix.

Because of the non-unitary dynamics implied in the Eqs.~(\ref{eq:ito_unnorm})-(\ref{eq:ito_unnorm2}), the norm of
the stochastically evolved bath density matrices are not conserved along each path of noise realization. 
The norm conservations only emerge after the noise averaging.  
This is a common source of numerical instabilities one encounters when directly simulating the simple stochastic dynamics presented so far.  

The norm fluctuations of $\tilde \rho_B(t)$ can be suppressed by modifying the stochastic differential equations above to read,
\begin{eqnarray}
\label{eq:ito}
& & d\tilde\rho_s = -i dt \left[\hat{H}_s, \tilde\rho_s\right] 
\mp i dt \sum_{\alpha} \left[ A^{\alpha},  \tilde\rho_s \right]_{\mp} \mathcal{B}^{\bar \alpha}(t)  \\
& & \, \, -\frac{i}{\sqrt 2} \sum_{\alpha} A^{\alpha}\tilde\rho_s dW_{\alpha}  + \frac{i}{\sqrt 2}\sum_{\alpha} \tilde\rho_s A^{\alpha} dV_{\alpha} , \nonumber \\
\label{eq:ito2}
& & d\tilde\rho_B = -i dt \left[\hat{H}_B, \tilde\rho_B\right] 
+\frac{1}{\sqrt 2} \sum_{\alpha} dW^*_{\bar \alpha} \left( B^{\alpha} \mp \mathcal B^\alpha(t) \right) \tilde \rho_B \nonumber \\
& & \, \, + \frac{1}{\sqrt 2}\sum_{\alpha} dV^*_{\bar \alpha} \tilde \rho_B  \left( B^\alpha - \mathcal B^\alpha (t) \right).
\end{eqnarray}
where additional stochastic fields
\begin{eqnarray}\label{eq:bfield}
\mathcal{B}^{\alpha} (t) = \left\{\begin{array}{ll} 
\sum_k g_k \text{Tr}_B \left\{\tilde\rho_B(t) b^{\alpha}_k  \right\}, & \text{complex-valued} \\
\alpha  \sum_k g_k \text{Tr}_B \left\{\tilde\rho_B(t) b^{\alpha}_k  \right\}, & \text{Grassmann-valued},
\end{array}\right.
\end{eqnarray}
is introduced to ensure $\text{Tr}_B
\tilde \rho_B(t)$ is conserved along each noise path.  In Eqs.~(\ref{eq:ito})-(\ref{eq:ito2}),  the top/bottom sign in the symbols
$\pm$ ($\mp$) refers to complex-valued/Grassmann-valued noises, respectively.
Similarly, $[\cdot,\cdot]_{\mp}$ refers to commutator (complex-valued noise) and anti-commutator (Grassmann-valued noise) in Eq.~(\ref{eq:ito}).
After these modifications, the exact reduced density matrix of the system is given by $\hat \rho_s(t) = \overline{\tilde \rho_s(t)}$.  

By introducting $\mathcal{B}^\alpha(t)$, we directly
incorporate the bath's response to the random noise in system's
dynamical equation.
Equation (\ref{eq:ito}) and the determination of $\mathcal{B}^\alpha(t)$ constitute the foundation of open system dynamics in the stochastic framework.  In addition to being a methodology of open quantum systems,
it should be clear that the present framework allows one to calculate explicitly bath operator involved quantities of interest\cite{Yan:2014cc}.

\subsection{Bath-Induced Stochastic Fields}\label{sec:bfield}
From Eq.~(\ref{eq:bfield}), it is clear that $\mathcal{B}^\alpha(t)$ can be obtained by formally integrating Eq.~(\ref{eq:ito2}).
For simplicity, we take $dV_\alpha = dV$ and similarly for $dW_{\alpha}=dW$.  This simplification does not compromise
the generality of the results presented below and applies to the common case of the spin boson like model when the 
system-bath interacting Hamiltonian
is given by $H_{\text{int}} = \sigma^z_0 \sum_{k>0} g_k (b^+_k + b_k)$ with $\mathcal{B}(t) = \mathcal{B}^+(t)+
\mathcal{B}^-(t)$.

We first consider the Gaussian baths composed of non-interacting bosons or fermions. 
The equations of motion for the creation and annihilation operators for individual modes read
\begin{eqnarray}
\label{eq:boson_mode1}
d\langle b^{\dag}_{k} \rangle & = & 
i\omega_{k} \langle b^{\dag}_{k} \rangle dt
+ \frac{1}{\sqrt 2} g_{k} dW^* \mathcal{G}^{+-}_k  
+ \frac{1}{\sqrt 2} g_{k} dV^* \mathcal{G}^{-+}_k,\\
d \langle b_{k} \rangle & = & -i \omega_{k} \langle b_{k} \rangle dt
+\frac{1}{\sqrt 2} g_{k} dW^* \mathcal{G}^{-+}_k 
\pm \frac{1}{\sqrt 2} g_{k} dV^* \mathcal{G}^{+-}_k, 
\label{eq:boson_mode2}
\end{eqnarray}
where the top (bottom) sign of $\pm$ should be used when complex-valued (Grassmann-valued) noises are adopted.
The expectation values in Eqs.~(\ref{eq:boson_mode1})-(\ref{eq:boson_mode2}) are taken with respect to $\tilde\rho_k(t)$.
The generalized cumulants are defined as
\begin{eqnarray}\label{eq:2ndcum}
\mathcal{G}^{\alpha_1 \alpha_2}_k = \langle b^{\alpha_1}_k b^{\alpha_2}_k \rangle \mp \langle b^{\alpha_1}_k \rangle  \langle b^{\alpha_2}_k \rangle.
\end{eqnarray}
For bosonic and fermionic bath models, it is straightforward to show that the time derivatives
of $\mathcal{G}^{\alpha_1 \alpha_2}_k$ vanish exactly.  Hence, the second order cumulants are determined by the thermal equilibrium conditions
of the initial states.  Immediately, one can identify the relevant quantity $\mathcal{G}^{+-}_k = n_{B/F}(\omega_k)$ 
representing either the Bose-Einstein or Fermi-Dirac distribution depending on whether it is a bosonic or fermionic mode.

Replacing the second order cumulants in Eqs.~(\ref{eq:boson_mode1})-(\ref{eq:boson_mode2}) with an appropriate thermal distribution, 
one can derive a closed form expression 
\begin{eqnarray}\label{eq:gaussbfield}
\mathcal{B}(t) 
& = & 
\frac{1}{\sqrt{2}}\int^t_0 dW^*_s \, \alpha(t-s)+\frac{1}{\sqrt{2}} \int^t_0 dV^*_s \, \alpha^*(t-s), 
\end{eqnarray} 
where 
\begin{eqnarray}
\alpha(t) = \left\{\begin{array}{ll}
\sum_k \vert g_k \vert^2 \left(\cos(\omega_k t) \coth(\beta\omega_k/2) - i \sin(\omega_k t)\right), & \text{bosonic bath} \\
\sum_k \vert g_k \vert^2 \left(\cos(\omega_k t) - i \sin(\omega_k t)\tanh(\beta\omega_k/2) \right), & \text{fermionic bath} \\
\end{array} \right.
\end{eqnarray}
stands for the corresponding two-time bath correlation functions. For the Gaussian bath models, Eqs.~(\ref{eq:ito}) and (\ref{eq:gaussbfield})
together provide an exact account of the reduced system dynamics.  

Next we illustrate the treatment of non-Gaussian bath models within the present framework with the spin bath as an example.  
In the rest of this section, the analysis only applies to the complex-valued noises.
The determination of $\mathcal{B}(t)$ still follows the same procedure described at the beginning of this section 
up to Eqs.~(\ref{eq:boson_mode1})-(\ref{eq:boson_mode2}).  The deviations appear when one tries to compute the time derivatives of the second cumulants.
Common to all non-Gaussian bath models, the second cumulants are not time invariant.  Instead, by iteratively using
Eq.~(\ref{eq:ito2}), the second and higher order 
cumulants can be shown to obey the following general equation,
\begin{eqnarray}\label{eq:gencum}
d \mathcal{G}^{\pmb \alpha}_k & = &
 i dt \vert \pmb{\alpha} \vert \omega_k \mathcal{G}^{\pmb \alpha}_{k}
+ \frac{1}{\sqrt{2}} dW^*_s \left(\mathcal{G}^{[\pmb{\alpha},+]}_{k}+\mathcal{G}^{[\pmb{\alpha},-]}_{k}\right) \nonumber \\
& & + \frac{1}{\sqrt{2}}dV^*_s \left(\mathcal{G}^{[+,\pmb{\alpha}]}_{k}+\mathcal{G}^{[-,\pmb{\alpha}]}_{k}\right),
\end{eqnarray}
where $\pmb \alpha = (\alpha_1, \alpha_2 \dots \alpha_n)$ specifies a sequence of raising and lowering spin operators that constitute this
particular $n$-th order cumulant and $\vert  \pmb\alpha \vert = \sum_i \alpha_i$ with $\alpha_i = \pm 1$ depending on whether it refers
to a raising (+) or lowering (-) operator, respectively.   The time evolution of these bath cumulants form a simple hierarchical structure with
an $n$-th order cumulant influenced directly by the $n+1$-th order cumulants according to the equation above where
we use $[\pmb \alpha, \pm] \equiv (\alpha_1 , \dots \alpha_n, \pm)$ to denote an $n+1$-th cumulant obtained by appending
a spin operator to $\pmb \alpha$.  A similar definition is implied for $[\pm, \pmb \alpha]$.  More specifically, 
these cumulants are defined via an inductive relation that we explicitly demonstrate with an example to obtain a third-order cumulant starting from a second-order one given in Eq.~(\ref{eq:2ndcum}),
\begin{eqnarray}
\mathcal{G}^{[(\alpha_1,\alpha_2),\pm]}_k = \langle b^{\alpha_1} b^{\alpha_2} (b^\pm - \langle b^\pm \rangle) \rangle + 
\langle b^{\alpha_1}(b^\pm - \langle b^\pm \rangle) \rangle \langle b^{\alpha_2}\rangle +
\langle b^{\alpha_1} \rangle \langle b^{\alpha_2} (b^\pm - \langle b^\pm \rangle) \rangle.
\end{eqnarray} 
The key step in this inductive procedure is to insert an operator identity $b^\pm - \langle b^\pm \rangle$ at the end of each expectation bracket 
defining the $n$-th cumulant. If a term is composed of m expectation brackets, then this insertion should apply to one bracket at a time and generate
m terms for the $n+1$-th cumulant.
Similarly, we get $\mathcal{G}^{[\pm,\pmb \alpha]}$ by inserting the same operator identity to the beginning of each expectation bracket of 
$\mathcal{G}^{\pmb \alpha}_k$.

For the spin bath, these higher order cumulants do not vanish and persist up to all orders.
In any calculations, one should certainly truncate the cumulants at a specific order by imposing the time invariance, 
$G^{\pmb \alpha}_k(t) = G^{\pmb \alpha}_k(0)$ and evaluate the lower order cumulants by recursively integrating
Eq.~(\ref{eq:gencum}).  Through this simple prescription, one derives
\begin{eqnarray}\label{eq:generalfield}
\mathcal B (t) 
& = & \frac{1}{\sqrt{2}}\int^t_0 dW^*_s \Phi_{2,1}(t,s) + 
\frac{1}{\sqrt{2}}\int^t_0 dV^*_s \Phi_{2,2}(t,s)  +  \nonumber \\
& & \left(\frac{1}{\sqrt 2}\right)^3 \int^t_0 \int^{s_1}_0 \int^{s_2}_0 dW^*_{s_1} dW^*_{s_2} dW^*_{s_3} \Phi_{4,1}(t,s_1,s_2,s_3) + \dots 
+\nonumber \\
& &  \left(\frac{1}{\sqrt 2}\right)^3 \int^t_0 \int^{s_1}_0 \int^{s_2}_0 dV^*_{s_1} dV^*_{s_2} dV^*_{s_3} \Phi_{4,8}(t,s_1,s_2,s_3) + \dots ,
\end{eqnarray}
where $\Phi_{n,m}(t_1, \dots, t_n)$ stands for an $n$-time bath correlation functions with the subscript $m$
used to distinguish the $2^{n-1}$ $n$-time correlation functions appearing in the stochastic integrations 
(each involves a unique sequence of noise variables) in Eq.~(\ref{eq:generalfield}).  Comparing to  Eq.(\ref{eq:gaussbfield}),one can identify $\Phi_{2,1}(t,s)$ and
$\Phi_{2,2}(t,s)$ with $\alpha(t-s)$ and $\alpha^*(t-s)$, respectively.
Note this derivation assumes the odd-time correlation functions vanish with respect to the initial thermal equilibrium state.

\subsection{Stochastic Liouville Equation}\label{sec:stochsummary}
  
At this point, we briefly summarize the unified stochastic formalism.
Once $\mathcal{B}(t)$ is fully determined, Eq.~(\ref{eq:ito}) can be presented in
a simple form, the stochastic Liouville equaiton (SLE),
\begin{eqnarray}\label{eq:sleq}
& & \frac{d \tilde\rho_s}{dt} = -i \left[\hat{H}_s, \tilde\rho_s\right]  
\mp i  A \tilde\rho_s(t)\xi(t)
 + i \tilde\rho_s(t)A\eta(t),\nonumber \\
\end{eqnarray}
where the newly defined color noises are
\begin{eqnarray}\label{eq:defcolornoise}
\xi(t)  & =  &  \mathcal{B}(t) \pm \frac{1}{\sqrt 2} \mu(t)  \nonumber \\ 
\eta(t) & =  &  \mathcal{B}(t) + \frac{1}{\sqrt 2} \nu(t). 
\end{eqnarray}
In these equations above, the top/bottom signs are associated with complex-valued/Grassmann-valued noises, respectively.

In the cases of bosonic baths, 
$\mathcal{B}(t)$ is given by Eq.~(\ref{eq:gaussbfield}) and driven by the complex-valued noises.
The color noises then are fully characterized by the statistical properties,
\begin{eqnarray}\label{eq:xi_corr}
\overline{ \xi(t) \xi(t') }& = & \alpha(\vert t-t' \vert), 
\nonumber \\
\overline{ \xi(t) \eta(t')} & = & \alpha(t'-t), 
\nonumber \\
\overline{\eta(t) \eta(t')}& = & \alpha^*(\vert t-t' \vert).
\end{eqnarray}
 Several stochastic simulation algorithms have been proposed to solve the SLE with the Gaussian noises.

On the other hand, all previous efforts in the stochastic formulations of fermionic bath end up with
derivations of either master equations\cite{Zhao_Yu_pra2012,Chen_You_pra2013} or 
hierarchical\cite{Suess:2015iu,Suess:2014gz} type of coupled equations. 
The stochastic framework has simply served as a mean to derive deterministic equations for numerical simulations.  
The lack of direct stochastic algorithm is due to the numerical 
difficulty to model Grassmann numbers.  In this study, we support the view that Grassmann numbers are simply ``formal bookkeeping devices" to help formulate the fermionic path integrals and the formal stochastic equations of motion with Gaussian properties. Hence, it is critical to formally eliminate the Grassmann number and the associated stochastic processes, which will be demonstrated in Sec.~\ref{sec:spin} with numerical 
illustrations in App.~\ref{app:num}.

Going beyond the Gaussian baths, $\mathcal{B}(t)$ is given
by Eq.~(\ref{eq:generalfield}) which involves multiple-time stochastic integrals. Formally, one can still
use the same definition of noises, Eq.~(\ref{eq:defcolornoise}), and the SLE
still prescribes the exact dynamics for the reduced density matrix.
The primary factor distinguishing from the Gaussian baths is the statistical characterization of the noises.
Higher order statistics are no longer trivial for non-Gaussian processes, and they 
are determined by the multi-time correlation
functions $\Phi_{nm}(t,t_1,\dots t_{n-1})$ in Eq.~(\ref{eq:generalfield}).  For instance, when the fourth
order correlation functions are included in the definition of $\mathcal{B}(t)$, additional statistical 
conditions such as $\overline{\xi(t_1) \xi(t_2) \xi(t_3) \xi(t_4)}$
would have to be imposed and related to $\{\Phi_{4m}(t_1,t_2,t_3,t_4),\Phi_{2m}(t_i,t_j)\}$ to fully specify this noise.  
Since constructing a purely stochastic method to simulate Gaussian processes is already a non-trivial task, simulating non-Gaussian random processes is an even tougher goal.

In the subsequent discussions, we should devise deterministic numerical methods based on the SLE, Eq.~(\ref{eq:sleq}),
by formally averaging out the noises. We name the proposed methods in Sec.~\ref{sec:spectral} - Sec.~\ref{sec:heom} collectively
as the generalized hierarchical equations (GHE) in this work.
Besides the GHE to be presented, we note
sophisticated hybrid algorithms\cite{Moix:2013jb,Zhou:2007fx} could also be constructed to combine advantages of both stochastic and deterministic 
approaches.  We shall leave these potential extensions in a future study.

\section{Solution I: Spectral Expansion of Stochastic Processes}\label{sec:spectral}
In this section, we consider a bosonic Gaussian bath.
We note Eq.~(\ref{eq:xi_corr}) encodes the full dissipative effects induced by a bath in the stochastic formalism.
The microscopic details, such as Eq.~(\ref{eq:defcolornoise}),
of the noise variables become secondary concerns.  
This observation allows us to substitute any pairs of correlated color
noises that satisfy Eq.~(\ref{eq:xi_corr}) as these statistical conditions alone 
do not fully specify the noises present in Eq.~(\ref{eq:sleq}). 
In other words, more than one set of noises can generate identical quantum dissipative dynamics as long as they all satisfy Eq.~(\ref{eq:xi_corr}) but may differ in other
unspecified statistics such as $\overline{\xi(t) \xi^*(t')}$ and $\overline{\eta(t) \eta^*(t')}$ etc.  
This flexibility with the choice\cite{qiang2016} of stochastic processes provides opportunities to fine-tune performances of nuemrical algorithms. 

We propose the following decomposition\cite{young_prl2013} of the noise variables
\begin{eqnarray}\label{eq:corr-stoch}
\xi(t) & = & \sum_{k} 
\left(\sqrt{\lambda^{R}_k} \psi_k(t) x_k + \sqrt{i \lambda^{I}_k} \phi_k(t) y_k + \sqrt{\lambda_k/2} \chi_k(t) z_k\right),   
\nonumber \\
\eta(t) & = & \sum_{k} \left(\sqrt{\lambda^{R}_k} \psi_k(t) x'_k - \sqrt{-i \lambda^{I}_k} \phi_k(t) y'_k + 
\sqrt{\lambda_k/2} \chi^*_k(t) z^*_k \right),   
\end{eqnarray}
where ($x_k$,$y_k$) and ($x'_k$, $y'_k$)
are independent and real-valued normal variables with mean $0$ and variance $1$ while $z_k=z^r_k + i z^i_k$ are similarly defined but complex-valued. The other unspecified
functions are obtained from the
spectral expansion of the correlation functions,
\begin{eqnarray}
\alpha(t-t') & = & \sum_k \lambda_k \chi^*_k(t) \chi_k(t'), \nonumber \\
\alpha_R(\vert t-t' \vert) & = & \sum_k \lambda^R_k \psi_k(t) \psi_k(t'), \nonumber \\
\alpha_I(\vert t- t' \vert) & = & \sum_k \lambda^I_k \phi_k(t) \phi_k(t'),
\end{eqnarray}
where $\alpha(t-t') = \alpha_R(t-t') + i \alpha_I(t-t')$, the functions $\chi_k(t)$ (complex-valued in general), 
$\psi_k(t)$ (real-valued) and $\phi_k(t)$ (real-valued) form independent sets of orthonormal basis function over time domain $[0,T]$ for the simulation.
The spectral components can be determined explicitly by solving
\begin{eqnarray}
\int^T_0 ds\, \alpha(t-s) \chi_k(s) = \lambda_k \chi_k(t),
\end{eqnarray}
and similar integral equations will yield other sets of basis functions and the associated eigenvalues.
The newly defined noises in Eq.~(\ref{eq:corr-stoch}) can be shown to reproduce the two-time statistics given in
Eq.~(\ref{eq:xi_corr}).

A crucial assumption of the spectral expansion above is that correlation functions should be 
positive semi-definite. This could be a concern with the quantum correlation functions
in the low temperature regime.  However, this problem can be addressed by modifying the
Hamiltonian and re-define the correlation functions in order to shift the spectral values by a large
constant to avoid negative eigenvalues.

We re-label the newly introduced random variables $g_k \in \{x_k, y_k, x'_k, y'_k, z^r_k, z^i_k\}$ and 
expand the reduced density matrix by\cite{young_prl2013} 
\begin{eqnarray}\label{eq:rho-stochexp}
\tilde\rho_s(t) & = & \sum_{\mathbf m} \sigma_{\mathbf m}(t) \Phi_{\mathbf m}(\mathbf g), \nonumber \\
& = & \sum_{\mathbf m} \sigma_{\mathbf m}(t) \mathcal H_{m_1}(g_1) \dots \mathcal H_{m_s}(g_s), 
\end{eqnarray}
where the function $\Phi_{\mathbf m}(\mathbf g)$ with $\mathbf{g}=(g_1,\dots g_s)$ is explicitly defined in the second line.
The $m$-th Hermite polynomial $\mathcal H_m(g)$ takes argument of random variables $g$.  The total number of random variables
$g_k$ is given by $s$. 
Every auxiliary density matrices $\sigma_\mathbf{m}(t)$ directly contributes to the determination of $\tilde\rho_s(t)$. 

Substituting Eqs.~(\ref{eq:corr-stoch}) and (\ref{eq:rho-stochexp}) into Eq.~(\ref{eq:sleq}) and 
average over all random variables $g_k$, one obtains a set of coupled equations for the density matrices,
\begin{eqnarray}
\partial_t \sigma_{\mathbf m} &= & -i[H_s, \sigma_{\mathbf m}] + i \sum_{k,\mathbf n} A \sigma_{\mathbf n} 
\theta_k(t) G^{k}_{\mathbf m \mathbf n} - i \sum_{k,\mathbf n} 
\sigma_{\mathbf n}A \theta^\prime_k(t) G^{k}_{\mathbf m \mathbf n},
\end{eqnarray}
where
$\theta_k(t) \in \{ \sqrt{\lambda_k^R}\psi_k(t), \sqrt{i \lambda^{I}_k} \phi_k(t), \sqrt{\lambda_k/2} \chi_k(t),
i\sqrt{\lambda_k/2} \chi_k(t)\}$, the
components of $\xi(t)$ in Eq.~(\ref{eq:corr-stoch}), and similarly $\theta^\prime_k(t)$ correspond to 
the components of $\eta(t)$, respectively.
In the above equation, $G^{k}_{\mathbf m \mathbf n}$ is defined by
\begin{eqnarray}
G^{k}_{\mathbf m \mathbf n} & = & 
\overline{\Phi_{\mathbf m}(\mathbf g) g_k \Phi_{\mathbf n}(\mathbf g)}\big/ \overline{\Phi_{\mathbf m}(\mathbf g)\Phi_{\mathbf m}(\mathbf g)}, 
\end{eqnarray}
where these averages can be done analytically by exploiting the properties of the Hermite polynomials and Gaussian integrals\cite{young_prl2013}.

Finally, the exact reduced density matrix is obtained after averaging out random variables in 
Eq.~(\ref{eq:rho-stochexp}), which can be done by invoking the Gaussian integral identities.
The present approach introduces an efficient decomposition of the noise variables and provides an alternative coupling structure
for a system of differential equations
than the standard HEOM in solving open quantum dynamics.
Unfortunately, the present method is not easily generalizable to accommodate the non-Gaussian 
processes.  

\section{Solution II: Generalized Hierarchical Equation of Motion}\label{sec:heom}
We next present another approach, starting from Eq.~(\ref{eq:sleq}) again, that yields 
deterministic equations and more easily to accommodate non-Gaussian bath models.  
This time we utilize Eq.~(\ref{eq:defcolornoise}) as the definitions for the noises, $\xi(t)$ and $\eta(t)$.
Following the basic procedure of Ref.~\onlinecite{Zhou:2007fx}, 
we average over the noises in Eq.~(\ref{eq:sleq}) to get
\begin{eqnarray}\label{eq:first-heom}
& & \frac{d\rho_s}{dt}  =  -i \left[\hat{H}_s, \rho_s\right] 
- i \left[ A, \overline{\tilde \rho_s \mathcal{B}} \right], 
\end{eqnarray}
The noise averages yield an auxiliary density matrix (ADM), 
$\overline{\tilde\rho_s\xi(t)} = \overline{\rho_s\eta(t)} = \overline{\tilde\rho_s \mathcal{B}(t)}$, by Eq.~(\ref{eq:defcolornoise}).  
Working out the equation of motion for the ADM, one is then required to define additional ADMs and solve their dynamics too. 
In this way, a hierarchy of equation of motions for ADMs develops with the general structure
\begin{eqnarray}\label{eq:genheom}
d_t[\overline{\tilde \rho_s \mathcal{B}^m}] = \overline{ d_t [\tilde\rho_s] \mathcal{B}^m} + \overline{\tilde \rho_s d_t \mathcal{B}^m} + 
\overline{d_t \tilde \rho_s d_t \mathcal B^m}.
\end{eqnarray}
The time derivatives of $\tilde\rho_s$ and $\mathcal{B}$ are given by Eqs.~(\ref{eq:ito}) and (\ref{eq:generalfield}), respectively.
If we group the ADM's into a hierarchical tier structure according to the exponent $m$ of $\mathcal{B}^m$, 
then it would be clear soon that the first term of the RHS of Eq.~(\ref{eq:genheom})
couples the present ADM to ones in the ($m+1$)-th tier, and the last term couples the present ADM to others in the ($m-1$)-th tier.  
  
In the rest of this section, we should materialize these ideas by formulating generalized HEOMs in detail.
We separately consider the cases of complex-valued noises (for bosonic and non-Gaussian bath models)
and Grassmann-valued noises (for fermionic bath models).

\subsection{Complex-Valued Noise}\label{sec:complex}
Following a recently proposed scheme, we introduce a complete 
set of orthonormal functions $\{\phi_j(t)\}$ and express all
the multi-time correlation functions in Eq.~(\ref{eq:generalfield}) as
\begin{eqnarray}
\Phi_{n+1,m}(t,t_1,\dots,t_n) = \sum_{\pmb j} \chi^{n+1,m}_{\pmb j} \phi_{j_1}(t-t_1) \cdots \phi_{j_n}(t_n-t_1),
\end{eqnarray}
where $\mathbf j = (j_1, \dots j_n)$.  Due to the completeness, one can also cast the derivatives of
the basis functions in the form,
\begin{eqnarray}
\frac{d}{dt} \phi_j(t) = \sum_{j'} \eta_{jj'} \phi_{j'}(t).  
\end{eqnarray}
Next we define cumulant matrices 
\begin{eqnarray}\label{eq:adef}
\mathbf{A}_n = \left[ \begin{array}{l}
a^n_{1 \mathbf{j}_1} \cdots a^n_{1 \mathbf{j}_k}, 0, \dots \\
\vdots \\
a^n_{2^{n} \mathbf{l}_1} \cdots  a^n_{2^{n} \mathbf{l}_{k^\prime}}, 0, \dots
\end{array}\right],
\end{eqnarray}
where each composed of $2^{n}$ row vectors with indefinite size.
For instance, $A_1$ has two row vectors while $A_2$ has four row vectors etc. The $m$-th row vector of matrix $A_n$ contains matrix elements denoted by $(a^n_{m \mathbf{j}_1}, a^n_{m \mathbf{j}_2} , \dots a^n_{m \mathbf{j}_k})$.  Each of this matrix element can be
further interpreted by
\begin{eqnarray}
a^n_{m \mathbf{j}}(t) \equiv \left(\frac{1}{\sqrt{2}}\right)^{n} \int^t_0 \int^{s_1}_0 \dots \int^{s_{n-1}}_0 dU^*_{s_1} \dots dU^*_{s_{n}} \nonumber \\
\phi_{j_1}(s_1-t) \phi_{j_2}(s_2-t)\dots\phi_{j_{n}}(s_n-t),
\end{eqnarray}
where $dU_{s_j}$ can be either a $dW_{s_j}$ or $dV_{s_j}$ stochastic variable depending on index $m$.
With these new notations, the multi-time correlation functions in Eq.~(\ref{eq:generalfield}) can be concisely encoded by
\begin{eqnarray}\label{eq:bosonb}
\mathcal{B}(t) = \sum_{n,m, \mathbf j} \chi^{n+1,m}_{\mathbf j} a^n_{m \mathbf j}(t).
\end{eqnarray}
Now we introduce a set of ADM's  
\begin{eqnarray}\label{eq:boson-adm}
& & \rho^{\left[\mathbf A_1 \right] \left[ \mathbf A_2 \right] \left[ \mathbf A_3 \right]} \dots \equiv  \overline{ \tilde\rho_s(t) \prod_{n,m,k} a^n_{m \mathbf{j}_k}(t)},
\end{eqnarray}
which implies the noise average over a product of all non-zero elements of each matrix $\mathbf{A}_i$ with the
stochastically evolved reduced density matrix of the central spin.  The desired reduced density matrix would correspond to the ADM at the zero-th tier with all $\mathbf A_i$ being null.  Furthermore, the very first ADM we discuss in Eq.~(\ref{eq:first-heom}) can be cast as
\begin{eqnarray}
\overline{ \tilde\rho_s \mathcal{B} (t)} = \sum_{n,m,\pmb j} \chi^{n,m}_{\pmb j}
\rho^{ \dots \left[\mathbf{A}_n\right] \dots }(t),
\end{eqnarray}
where each ADM on the RHS of the equation carries only one non-trivial matrix element $a^{n}_{m,\pmb j}$ in
$\mathbf{A}_n$.
Finally, The hierarchical equations of motion for all ADMs can now be put in the following form,
\begin{eqnarray}\label{eq:gheom}
& & \partial_t \rho^{\left[ \mathbf A_1 \right] \left[ \mathbf A_2 \right] \left[ \mathbf A_3 \right] \cdots}  =  
-i \left[ H_s, \rho^{\left[ \mathbf A_1 \right] \left[ \mathbf A_2 \right] \left[ \mathbf A_3 \right] \cdots} \right] 
-i \sum_{n,m,\pmb j} \chi^{n+1,m}_{\pmb j} 
\left[ A, \rho^{\cdots \left[ \mathbf{A}_n + (m,\pmb j) \right] \cdots}  \right] \nonumber \\
& &  -i \sum_{n,m,\mathbf j} 
\phi_{j_1}(0) A \rho^{\cdots \left[ \mathbf{A}_{n-1} + (m',\pmb{j}_1) \right]\left[ \mathbf{A}_n - (m,\pmb j) \right] \cdots} 
 -i \sum_{n,m,\mathbf j} 
\phi_{j_1}(0) \rho^{\cdots \left[ \mathbf{A}_{n-1} + (m',\pmb{j}_1) \right]\left[ \mathbf{A}_n - (m,\pmb j) \right] \cdots} A \nonumber \\
& & + \sum_{n,m,\mathbf{j}\mathbf{j'}} \eta_{\mathbf{j} \mathbf{j'}}  
\rho^{\cdots \left[ a^{n}_{m \mathbf{j}} \rightarrow a^{n}_{m \mathbf{j'}} \right] \cdots}
\end{eqnarray}

This equation involves a few compact notations that we now explain.  We use $\left[ \mathbf{A}_n \pm (m,\pmb j) \right]$ to mean adding or removing
an element $a^{n}_{m \pmb j}$ to the $m$-th row.  We also use  $\left[ a^{n}_{m \mathbf{j}} \rightarrow a^{n}_{m \mathbf{j'}} \right]$ to denote
a replacement of an element in the $m$-th row of $\mathbf{A}_n$. On the second line, we specify an element in a lower matrix given by $(m',\mathbf{j}_1)$.
The variable $\mathbf{j}_1$ implies removing the first element of the $\mathbf{j}$ 
array and the associated index $m'$ is determined by removing the first stochastic integral in Eq.~(\ref{eq:adef}). 
We caution that there is no $\mathbf{A}_0$ matrix and such a term whenever arises should simply be ignored when interpreting the above equation. 
After the first term on the RHS of Eq.~(\ref{eq:gheom}), we only explicitly show the matrices $\mathbf{A}_n$ affected in each term of the equation.

This generalized HEOM structure reduces exactly to the recently proposed eHEOM\cite{Tang:2016gh} when only 
$\mathbf{A}_1$ cumulant matrix 
carries non-zero elements, i.e.
only the second cumulant expansion of an influence functional is taken into account. 
It is clear that the higher-order non-linear effects induced by the bath's $n$-time correlation functions will only appear earliest
at the ($n-1$)-th tier expansion.

\subsection{Grassmann-valued Noise}\label{sec:fermion}
Next we consider the fermionic bath models with Grassmann-valued noises. As discussed earlier, the Grassmann numbers
are essential to manifest the Gaussian properties of fermionic baths.
In this case, one can significantly simplify the generalized HEOM in the previous section.
First, the bath-induced stochastic field can still be decomposed in the form,
\begin{eqnarray}\label{eq:fermion-corrfn}
\mathcal{B}(t) = \sum_{j=1}^{K} \chi_{j} a_{j}(t).
\end{eqnarray}
where the indices $m$ and $n$ are suppressed when compared to Eq.~(\ref{eq:bosonb}).
This is because the Hamiltonian we consider in this study only allows fermion Gaussian bath model.
Each element $a_{j}$ are similarly defined
\begin{eqnarray}
a_{j}(t) \equiv \left(\frac{1}{\sqrt{2}}\right)\int^t_0 dU^*_{s} \phi_{j}(t-s),
\end{eqnarray}
where $dU = dW (dV)$ (Grassmann-valued) when $j \leq K/2$ (or $ > K/2$). 
Similar to the algebraic properties of fermionic operators, there are no higher powers of Grassmann numbers and
each element $a_{m \mathbf j}$ can only appear once.  This Pauli exclusion 
constraint allows us to simplify the representation of 
fermionic ADMs.
We may specify an m-th tier ADM by
\begin{eqnarray}\label{eq:fm-tier-defn}
\bar \rho^{m}_{\mathbf{n}}(t)  & = & \overline{ \tilde\rho_s(t) a^{n_1}_{1}(t) \cdots a^{n_{K}}_{K}(t)},
\nonumber \\
\end{eqnarray}
where $\mathbf{n}=(n_1, \dots n_{K})$ with $n_i = 0$ or $1$.  
In this simplified representation, instead of specifying the non-zero elements as in Eq.~(\ref{eq:boson-adm}),
we layout all elements $a_{mj}$ in an ordered fashion and employ the binary index $n_i$ to denote which basis
functions contribute to a particular ADM.  The tier level of an ADM is determined by the number of basis
participating functions, i.e.
$m= \sum_i n_i$.

Following the general procedure outlined in Eq.~(\ref{eq:genheom}), the m-th tier HEOM reads,
\begin{eqnarray}\label{eq:fm-tier-heom}
 \frac{d \bar \rho^{m}_{\mathbf n}}{dt} & = & -i \left[ \hat{H}_s,\rho^m_{\mathbf n} \right]  
+i \sum_{j} \left( \chi_j A \bar\rho^{m+1}_{\mathbf n + \mathbf 1_{j}} (-1)^{\vert \mathbf n \vert_j} + 
\chi_j \bar\rho^{m+1}_{\mathbf n + \mathbf 1_{j}} A (-1)^{\vert \mathbf n \vert_j} \right)(1-n_j) 
\nonumber \\
& & 
+i\sum_{j} \phi_{j}(0) A \rho^{m-1}_{\mathbf n - \mathbf 1_j}(-1)^{\vert \mathbf n \vert_j}n_j
\nonumber \\
& & 
+i\sum_{j} \phi_j(0) \rho^{m-1}_{\mathbf n - \mathbf 1_j} A (-1)^{\vert \mathbf n \vert_j}n_j
\nonumber \\
 & & +
\sum_{j,j'} \eta_{jj'} \rho^{m}_{\mathbf n_{j,j'}} (-1)^{\vert \mathbf n \vert_j+\vert \mathbf n \vert_{j'}}
n_j (1-n_{j'}),
\end{eqnarray}
where $\vert \mathbf n \vert_j = \sum_{i=0}^{j-1} n_j$ and $\mathbf n_{j,j'}$ implies setting $n_j=0$ and $n_{j'}=1$. 
In the above equation, $\mathbf 1_j$ is a vector of zero's except an one at the $j$-th component.
The factor such as $n_j$ and $(1-n_j)$ are present to enforce the Pauli exclusion principle associated with
the fermions.

The structure of fermionic HEOM certainly resembles that of the bosonic case. However, a few distinctions are worth emphasized. First, it is just the Gaussian bath result including only $\mathbf{A}_1$ block matrix when compared to the results in sec.~\ref{sec:complex}.
Secondly, 
the fermionic HEOM truncates exactly at some finite number of tiers due to the constraint on the array of binary indices, $n_i$. 
Extra negative signs arise from the permutations to shift the underlying Grassmann-valued stochastic variables 
from their respective positions in 
Eq.~(\ref{eq:fm-tier-defn}) to the left end of the sequence. 

\section{Spin Bath: Dual-Fermion Transformation}\label{sec:spin}

We dedicate an entire section to discuss the spin bath from   
two distinct perspectives. If one formulates the SLE for a spin bath model in terms of complex-valued noises, then
the bath induced stochastic field $\mathcal{B}(t)$ is given by Eq.~(\ref{eq:generalfield}).  In this way, the spin 
bath is a specific example of non-Gaussian bath models. The generalized HEOM formulated earlier can be directly
applied in this case.  However, in any realistic computations, it is necessary to truncate the statistical characterization
of $\mathcal{B}(t)$ up to a finite order of multi-time correlation functions in Eq.~(\ref{eq:generalfield}).  While the method
is numerically exact, it is computationally prohibitive to calculate beyond the first few higher-order corrections.      
When the spin bath is large and can be considered as a finite-size approximation to a heat bath, one can show that the linear
response approximation\cite{makri99} often yields accurate results and a leading order correction should be sufficient whenever needed.  The relevance of this leading order correction for spin bath models 
will be investigated in the paper II\cite{hsieh_cao_jcp16}.

On the other hand, a spin bath composed of nuclear / electrons spins, as commonly studied in artificial nanostructures at ultralow temperature regime, can beahve very differently from a heat bath composed of non-interacting bosons.
There is no particular reason that the linear response and the first leading order correction should sufficiently account for quantum dissipations under all circumstances.  
In this scenario, it could be useful to map each spin mode onto a pair of coupled fermions.
The non-linear mapping allows us to efficiently capture the exact dynamics in an extended Gaussian bath model. 

\subsection{Dual-Fermion Representation}\label{sec:spin-dualf}
We consider the following transformation that maps each spin mode into two fermions via, 
\begin{eqnarray}
\sigma^x_k  =  (c^\dag_k - c_k)(d^\dag_k + d_k), \,\,\, 
\sigma^y_k  =  i(c^\dag_k + c_k)(d^\dag_k + d_k),\,\, \,
\sigma^z_k =  -2\left(c^\dag_k c_k - \frac{1}{2}\right),  
\end{eqnarray}
where the fermion operators satisfy the canonical anti-commutation relations.
One can verify the above mapping reproduces the correct quantum angular momentum commutation relations
with the $c$ fermion operators for each spin, while the presence of additional $d$ fermions 
makes the spin operators associated with different modes commute with each other.   
We now re-write the Hamiltonian as
\begin{eqnarray}\label{eq:dfH}
H 
  & = & H_s + \sum_{k>0} \omega_k \left(c^\dag_k c_k -\frac{1}{2}\right) \nonumber \\
  & & - \sum_{k>0} g_k \sigma^z_0 \left( d^\dag_k + d_k \right) \left( c^\dag_k - c_k \right).
\end{eqnarray}
The initial density matrix still maintains a factorized form in the dual-fermion representation,
\begin{eqnarray}
\hat \rho(0) = \hat \rho_s(0) \hat \rho_c^{eq}(\beta) \hat I_{N},
\end{eqnarray}
where $\hat \rho_c^{eq}(\beta)$ is the thermal equilibrium state of the $c$ fermions at the original inverse temperature $\beta$ of the spin bath and the $d$ fermions are in the maximally mixed state which is denotes by the identity matrix with dimension $N = 2^n$ where $n$ is the number of bath modes.  
A normalization constant is implied to associate with the $\hat I _N$ matrix.

According to the transformed Hamiltonian
in Eq.~(\ref{eq:dfH}), the system bath coupling now involves three-body interactions,
$\sigma^z_0 (d_k^\dag + d_k)(c_k^\dag-c_k) $. Furthermore, the two femrionic baths portrait a non-equilibrium setting
with $c$ fermionic bath inherits all physical properties of the original spin bath while $d$ fermionic bath is always 
initialized in the infinite-temperature limit regardless of the actual state of the spin bath.
We first take the system and the d fermions together as an
enlarged system and treat the c fermions collectively as a fermionic bath.  
We introduce the Grassmann noises to stochastically decouple the two subsystems,
\begin{eqnarray}
\label{eq:sd-ito}
d\tilde \rho_{sd} & = & -i dt \left[ \hat H_{s},\tilde \rho_{sd} \right] 
- i \sum_{k} \sigma^z_0 A_k \tilde \rho_{sd} d\mathcal{W}_k, \nonumber \\
& & \,
+ i \sum_k \tilde\rho_{sd} \sigma^z_0 A_k d\mathcal{V}_k \nonumber \\
\label{eq:sd-ito2}
d\tilde \rho_{k} & = & -i dt \left[ \hat H_{b,k}, \tilde \rho_k \right] + \frac{dW^*_k}{\sqrt 2} 
\left(B_k+\mathcal{B}_k\right) \tilde \rho_k \nonumber \\
& & \, +  \frac{dV^*_k}{\sqrt 2} \tilde \rho_k \left(B_k-\mathcal{B}_k\right),
\end{eqnarray}
where $B_k = g_k(c^\dag_k-c_k)$, $\mathcal{B}_k = g_k\text{Tr}\left\{(c^\dag_k+c_k)\tilde\rho_k(t)\right\}$,
$A_k = (d^\dag_k + d_k)$ and $H_{b,k} = \omega_k c^\dag_k c_k$. The density matrices $\tilde\rho_{sd}$
denotes the extended system including system spin and all $d$ fermions and $\tilde\rho_{k}$ denotes the individual 
c fermions with $k = 1, \dots, N_B$.  In Eq.~(\ref{eq:sd-ito}), the noises are defined by
\begin{eqnarray}
d\mathcal{W}_k &=& \frac{1}{\sqrt 2} dW_k - \mathcal{B}_k, \nonumber \\
d\mathcal{V}_k &=& \frac{1}{\sqrt 2} dV_k - \mathcal{B}_k,
\end{eqnarray}
where $dW_k$ and $dV_k$ are the standard Grassmann noises defined earlier.
The Eq.~(\ref{eq:sd-ito2}) clearly conserve the norm of $\tilde\rho_k(t)$ along each noise path, and we will
focus on Eq.~(\ref{eq:sd-ito}) and the stochastic fields, $\mathcal{B}_k(t)$.


Our main interest is just the system spin.  Hence, we trace out the d fermions in Eq.~(\ref{eq:sd-ito})
and get
\begin{eqnarray}\label{eq:dual-stoch1}
d\psi^{0}_0 & = & \text{Tr}_d \left\{ d\rho_{sd} \right\} \nonumber \\
& = & -i dt \left[ H_s, \psi^0_0\right] + i \sum_k \left[ \sigma^z_0, \psi^{1}_k\right] \frac{dX_k}{2} \nonumber \\
& & +i \sum_{k} \left\{\sigma^z_0, \psi^{1}_k \right\}\left(\frac{dY_k}{2} - \mathcal{B}_k \right).
\end{eqnarray}
The auxiliary objects, $\psi^n_\mathbf k$, appearing in Eq.~(\ref{eq:dual-stoch1}) are defined via
\begin{eqnarray}
\psi^{n}_\mathbf{k} = \text{(s)Tr}\left\{ A_{k_1} \cdots A_{k_n} \tilde \rho_{sd}\right\},
\end{eqnarray}
where $k_1 > \cdots > k_n$, (s)Tr either implies standard trace (n is even) or super-trace (n is odd),
and the new noises
\begin{eqnarray}
dX_k = \frac{dW_k + i dV_k}{\sqrt 2} \,\, \text{and} \,\, dY_k = \frac{dW_k -i dV_k}{\sqrt 2}.
\end{eqnarray}

A hierarchical structure is implied in Eq.~(\ref{eq:dual-stoch1}), so we derive
the equations of motions for the auxiliary objects, 
\begin{widetext}
\begin{eqnarray}\label{eq:dual-stoch2}
& & d\psi^{n}_\mathbf{k} = -i dt \left[H_s,\psi^n_\mathbf{k}\right]
+i \sum_{j \notin \mathbf{k}}  \left\{\sigma^z_0,
\psi^{n+1}_{\mathbf{k}+j}\right\} \frac{dX_k}{2}(-1)^{\vert \mathbf k \vert_{> j}} 
\, +i \sum_{j \in \mathbf{k}}  \left[\sigma^z_0,
\psi^{n-1}_{\mathbf{k}-j}\right]\frac{dX_k}{2} (-1)^{\vert \mathbf k \vert_{\geq j}}\nonumber \\
& & +i \sum_{j \notin \mathbf{k}}  \left[\sigma^z_0,
\psi^{n+1}_{\mathbf{k}+j}\right] \left(\frac{dY_k}{2} - \mathcal{B}_k \right)
(-1)^{\vert \mathbf k \vert_{> j}}
+i \sum_{j \in \mathbf{k}}  \left\{\sigma^z_0,
\psi^{n-1}_{\mathbf{k}-j}\right\}\left(\frac{dY_k}{2} - \mathcal{B}_k \right) 
(-1)^{\vert \mathbf k \vert_{\geq j}},
\end{eqnarray}
\end{widetext}
where $\vert \mathbf k \vert_{\geq j} \equiv \sum_{k_i \geq j} k_i$.

Since the spin bath model is mapped onto an effective fermionic problem, the uses of Grassmann noises,
Eq.~(\ref{eq:dual-stoch2}), will serve as a starting point to develop deterministic numerical methods
once the Grassmann noises are integrated out.  

\subsection{Dual-Fermion GHE}\label{sec:heom-spin}
To solve Eq.~(\ref{eq:dual-stoch2}), we first define the generalized ADMs
\begin{eqnarray}\label{eq:adm_sbath}
\rho^{n,m}_{\Omega, \mathbf j} =\overline{\psi^{n}_\mathbf{k} \mathcal{B}^{\alpha_n}_{k_n} \cdots 
\mathcal{B}^{\alpha_1}_{k_1} \left( \mathcal{B}^+_{j_m} \mathcal{B}^-_{j_m} 
\cdots \mathcal{B}^+_{j_1} \mathcal{B}^-_{j_1}\right)}
\end{eqnarray}
where $\Omega = (\mathbf k, \pmb{\alpha})$. 
The 2 index vectors $\mathbf k$ and $\mathbf j$ label pairs of coupled $c$ and $d$ fermionic modes; 
furthermore, the two index vectors are mutually exclusive in the sense a bath mode can appear  
in just one of the two vectors each time. In this case, the tier-structure of the ADMs are determined by $n+m$.
Same as the bosonic and fermionic bath results, the desired reduced density matrix is exactly given by the zero-th tier of ADMs.

Further notational details of Eq.~(\ref{eq:adm_sbath}) are explained now.
The vector $\pmb{\alpha}$ is to be paired with the vector $\mathbf k$ to characterize the first set of stochastic fields $\mathcal{B}^{\alpha}_k$ in Eq.~(\ref{eq:adm_sbath}).
More precisely, each $(k_i,\alpha_i = \pm)$ labels one of the two possible stochastic fields,
$\mathcal{B}^{+}_{k_i} = g_{k_i} \langle c^{\dag}_{k_i} \rangle$ and 
$\mathcal{B}^{-}_{k_i} = g_{k_i} \langle c_{k_i} \rangle$, associated with $k_i$-th c fermion.  
In dealing with bosonic, fermionic and non-Gaussian baths, we need to explicitly use bath's multi-time correlation
functions via 
$\mathcal{B} = \sum_{k} \left(\mathbf B^+_{k_i} \pm \mathbf{B}^+_{k_i}\right)$ when formulating the generalized HEOM 
approach.
In the present case, the stochastic decoupling we introduced in Sec.~\ref{sec:spin-dualf} dictates that  
each c fermion acts as a bath and equipped with its own set of stochastic fields as shown in 
Eq.~(\ref{eq:dual-stoch2}).
There is no need to expand the bath correlation functions in some orthonormal basis,
as each mode's correlation functions will be treated explicitly in a Fourier decomposition.

Repeat the same steps of the derivation as before, we obtain the generalized HEOM for the spin bath,
\begin{widetext}
\begin{eqnarray}\label{eq:dualfheom}
& & \partial_t \rho^{n,m}_{\Omega, \mathbf j} = -i[H_s, \rho^{n,m}_{\Omega, \mathbf j}] 
+ i \sum_{l \in \mathbf k} \alpha_l \omega_l \rho^{n,m}_{\Omega, \mathbf j} 
+ i \sum_{\substack{l \notin \mathbf k, l \notin \mathbf j \\  \gamma}}
\left[ A, \rho^{n+1,m}_{\Omega+(l,\gamma), j}\right] 
+i \sum_{l \in \mathbf k} \alpha_l \left\{ A, \rho^{n-1,m+1}_{\Omega-(l,\alpha_l),\mathbf j+\mathbf 1_l}\right\} \nonumber \\
& & -\frac{i}{2} \sum_{l \in \mathbf j, \gamma} \gamma g_l^2 (1-2n_F(\omega_l)) \left\{ A, \rho^{n+1,m-1}_{\Omega+(l,\bar\gamma),\mathbf j - \mathbf 1_l} \right\}
-\frac{i}{2} \sum_{l \in \mathbf k} g_l^2 (1-2n_F(\omega_{l})) \left[ A, \rho^{n-1,m}_{\Omega-(l,\alpha_l),\mathbf j} \right], \nonumber \\
& & +\frac{i}{2} \sum_{l \in \mathbf j, \gamma} g_l^2 \left[ A, \rho^{n+1,m-1}_{\Omega+(l,\gamma),\mathbf j - \mathbf 1_l}\right]
+\frac{i}{2} \sum_{l \in \mathbf k} \alpha_l g_l^2 \left\{A, \rho^{n-1,m}_{\Omega-(l,\alpha_l),\mathbf j}\right\} ,
\end{eqnarray}
\end{widetext}
where $\Omega \pm (l,\alpha_l)$ means a stochastic field $\mathcal{B}^{\alpha_l}_l$ is either added or removed from the vectors $\mathbf k$ and $\pmb{\alpha}$ and, similarly, $\mathbf j \pm \mathbf 1_l$ means an index $j=l$ is
either added or removed from $\mathbf j$.

The range of the index values $\mathbf k$ and $\mathbf j$ can be extremely large as we explicitly label each microscopic bath modes.  Due to the hierarchical structure and the way ADMs are defined, it becomes prohibitively
expensive to delve deep down the hierarchical tiers in many realistic calculations.  However, the situation
might not be as dire as it appears.  We already discuss how the present formulation is motivated by the 
physical spin based environment, such as a collection of nuclear spins in a solid. 
In such cases, the bath often possess some symmetries allowing simplifications.  
For instance, most nuclear spins will precess at the same Lamour frequency, and the coupling constant is often 
distance-dependent.  Hence, 
one can construct spatial ``symmetric shells" centered around the system spin in the 3-dimensional real space such that all bath spins inside a shell will more or less share the same frequency and system-bath coupling coefficient.
By exploiting this kind of symmetry arguments, one can combine many ADMs defined in Eq.~(\ref{eq:adm_sbath}) together to significantly reduce the complexity of the hierarchical structures.
For a perfectly symmetric bath (i.e. one frequency and one system-bath coupling term),
one can use the following compressed ADM,
\begin{eqnarray}\label{eq:dualfc}
\rho^{n;s} & = & \sum_{|\pmb\alpha|=s, |\mathbf k|=n} \overline{\psi^{n}_\mathbf{k} \mathcal{B}^{\alpha_n}_{k_n} \cdots 
\mathcal{B}^{\alpha_1}_{k_1} },
\end{eqnarray}
where the sum takes into account of all possible combinations of $n$ modes compatible with the requirement that $\sum_i \alpha_i = s$.

On the other hand, if one deals with a large spin bath described by an effective spectral density then treating
the spin bath as an anharmonic environment and usage of 
the generalize HEOM in Sec.~\ref{sec:complex} will be more appropriate. In fact, in the thermodynamical limit, 
the spin bath can be accurately approximated as a Gaussian bath and one only needs to invoke $\mathbf{A}_1$ block
matrix in most calculations.

\section{Concluding Remarks}\label{sec:last}
In summary, we advocate the present stochastic framework as a unified approach to extend the study of dissipative quantum dynamics beyond the standard bosonic bath models. 
We exploit the It\^{o} calculus rule to represent
any bilinear interaction between two quantum DOF as white noises.
Starting from Eqs.~(\ref{eq:ito})-(\ref{eq:ito2}), one can derive the SLE, Eq.~(\ref{eq:sleq}), with appropriate
statistical conditions, such as Eq.~(\ref{eq:xi_corr}), that the noises must satisfy. In the Gaussian bath models,
the required conditions only involve two-time statistics determined by the bath's correlation functions.
In the case of non-Gaussian bath models, the noises are further characterized by higher order statistics 
and the multi-time correlation functions.

We devise a family of GHE to solve the SLE with deterministic simulations. 
We consider two separate orthonormal basis expansions: (1)
spectral expansion and (2) generalized HEOM.
The spectral expansion, in Sec.~\ref{sec:spectral}, allows us to solve bosonic bath models
efficiently when bath's two-time correlations assume a simple spectral expansion.  This is often 
the case for correlation functions with a slow decay. 
The second approach, in Sec.~\ref{sec:heom}, generalizes the eHEOM method to handle multi-time correlation
functions in some arbitary set of orthnormal functions.  This generalization can provide numerically
exact simulations for non-Gaussian (including spin), fermionic and bosonic bath models with arbitray sepctral densities and
temperature regimes.   

Among the bath models, we extensively discuss the spin bath.
When a spin bath is characterized by a well-behaved spectral density\cite{makri99}, 
the generalized HEOM in Sec.~\ref{sec:heom}
serves as an efficent approach to simulate dissipative quantum dynamics in a non-Gaussian bath.
For situations requiring more than a few higher-order response functions,
such as baths composed of almost identical nuclear / electron spins,
an alternative approach is to first map the spin bath onto an enlarged Gaussian bath model of fermions via the dual-fermion representation and apply the dual-fermion GHE in Sec.~\ref{sec:heom-spin}.  
Numerical examples are illustrated in App.~\ref{app:num}.

\begin{acknowledgements}
C.H. acknolwedges support from the SUTD-MIT program.  J.C. is supported by NSF (grant no. CHE-1112825) and SMART.
\end{acknowledgements}

\appendix
\section{Stochastic Processes}\label{app:stochprocess}
We focus on the complex-valued stochastic processes in this appendix.  
Additional remarks on Grassmann noises will be made
in the following section.

The basic Wiener processes considered in this work is taken to be $W(t) = \int^t_0 ds \nu(s)$ where the
complex-valued noise has a mean $\overline{\nu(t)}=0$ and a variance $\overline{\nu(t)\nu^*(t')}=2\delta(t-t')$.
Take a uniform discretization of time domain, in each time interval $dt_i = t_{i}-t_{i-1}$, each white noise path
reduces to a sequence of normal random variables $\{\nu_1 \dots \nu_N\}$. Hence, at each time interval,
an identical normal distribution is given, 
\begin{eqnarray}
P_i(\nu,\nu^*) = \frac{\Delta t}{2\pi}e^{-\frac{\Delta t}{2}\vert \nu \vert^2}
\end{eqnarray} 
The variance is chosen to reproduce the Dirac delta function in the limit $\Delta t \rightarrow 0$.
Furthermore, the differential Wiener increments $dW(t)=\nu(t) dt$ satisfy $dW(t)dW^*(t) \sim dt$ as
required for Brownian motion. 
The averaging process, implied by the bar on top of stochastic variables, can now be explicitly defined
as
\begin{eqnarray}
\overline{f(\{\nu_i,\nu^*_i\})} 
= \prod_{i=1}^N \int d\nu_{i} d\nu^*_{i} P_i(\nu_i,\nu^*_i) f(\{\nu_i,\nu^*_i\}).
\end{eqnarray}

\section{Grassmann Numbers and Noises}\label{app:grassmann}
Grassmann numbers are algebraic constructs that anti-commute among themselves and with any fermionic operators.
Given any two Grassmann numbers, $x$ and $y$, and a fermionic operator, $\hat c$, they satisfy
\begin{eqnarray}
 xy = - yx \,\,\,\,\,\, \text{and} \,\,\,\,\,\, x \hat c = - \hat c x.  
\end{eqnarray}
Furthermore, the Grassmann numbers commute with the vacuum state $\ket{0}$ and, consequently, anti-commute with 
$\ket{1}=\hat c^\dag\ket{0}$.
Besides the fermionic operators, these numbers commute with everything else such
as the bosonic operators and spin Pauli matrices.  

Due to the anti-commutativity, there is no higher powers of Grassmann numbers, i.e. $x^2 = 0$.
For instance, a single-variate Grassmann function $F(x) = a + bx$ (all variables, a, b, and x, 
are Grassmann-valued) can only assume this finite Taylor-expanded form.
In general every Grassmann function can be decomposed into odd and even parity, $F(x) = F_{\text{even}}(x) + F_{\text{odd}}(x)$.
such that
\begin{eqnarray}
F_{\text{odd}}(x) G(x) & = & G(-x) F_{\text{odd}}(x), \nonumber \\
F_{\text{even}}(x) G(x) & = & G(x) F_{\text{even}}(x),
\end{eqnarray}
where $G(x)$ is another arbitrary function with no particular parity assumed.
The fermionic thermal equilibrium states are are even-parity Grassmann function
when represented in terms of the fermonic coherent states. This even-parity is preserved under linear driving with Grassmann-valued noises. 
This means all the Grassmann numbers will commute with the fermonic bath density matrices in our study.

Another relevant algebraic property for our study is
\begin{eqnarray}
\text{Tr}\left(x \hat \rho\right) & = &
x \, \text{sTr} \left( \hat \rho\right) \nonumber \\
& = & x \left(\bra{0} \hat \rho \ket{0} - \bra{1} \hat \rho \ket{1} \right),
\end{eqnarray}
where $x$ is a Grassmann number and sTr$\left\{ \cdot \right\}$ is often termed the super-trace.

Finally, we discuss Grassmann-valued white noises.
Similar to the discretized complex-valued white noises introduced earlier, 
we shall take the noise path as a continuum limit of a sequence of Grassmann numbers, $\{x_i\}\vert_{i=1}^N$. 
We will formally treat them as random numbers with respect to Grassmann Gaussians as probability distributions.
More precisely, the following integrals yield the desired first two moments (in analogy to the complex-valued normal random variables),
\begin{eqnarray}
\overline{x} & = & \frac{2}{\Delta t} \int dx^* dx e^{-\frac{\Delta t}{2} x x^*} x = 0, \nonumber \\
\overline{x x^*} & = & \frac{2}{\Delta t}\int dx^* dx e^{-\frac{\Delta t}{2} x x^*} xx^* = \frac{2}{\Delta t}
\end{eqnarray} 
where the Gaussians should be interpreted by the Taylor expansion: $e^{x x^\prime} = 1 + x x^\prime$.  In evaluating the integrals above, we recall 
the standard Grassmann calculus rule that
integration with respect to $x$ is equivalent to differentiation with respect to $x$. With these basic set-ups,
one can operationally formulate Grassmann noises in close analogy to the complex-valued cases.

\section{Relating Stochastic Formalism and Influence Function Theory}\label{app:inf-func}
The connection between the two formalisms is usually investigated by deriving the stochastic equations from the influence functional theory via the Hubbard-Stratonovich transformation\cite{Shao:2004et,Stockburger:2002em}.  Nevertheless, to advocate the stochastic
view of quantum dynamics as a rigorous foundation, we establish the connections in the reversed order.  We should restrict to the standard bosonic bath models, but extension should be obvious.
We first re-write Eq.~(\ref{eq:ito_unnorm}) as
\begin{eqnarray}
\label{eq:app_ito_unnorm}
& & \partial_t \tilde\rho_s  = 
-i H_L(t) \tilde \rho_s + i \tilde\rho_s H_R(t) \nonumber \\
& & \,\, -i \left[H_s + \frac{\mu^*(t)}{\sqrt{2}} A\right] \tilde\rho_s 
+i \tilde\rho_s \left[H_s -i \frac{\nu^*(t)}{\sqrt{2}}A\right],
\end{eqnarray}
where the system-bath interaction is given by Eq.~(\ref{eq:genH}).
In this revised form, it is immediately clear that
\begin{eqnarray}\label{eq:appc-rdm}
& & \tilde\rho_s(t) = \\
& & \, T_+\exp\left(-i \int^t_0 ds H_L(s)\right) \rho_s(0) T_-\exp\left(+i \int^t_0 ds H_R(s)\right),\nonumber 
\end{eqnarray}
where $T_\pm$ is the time-ordering (+) and anti-time-ordering (-) operator.
By inserting a complete set of basis $\{\ket{\alpha}\}$ at each time slice, Eq.~(\ref{eq:appc-rdm})
can be put in the form,
\begin{eqnarray}\label{eq:appc-rdm2}
& & \tilde\rho_s(\pmb{\alpha}_t; t) = \\
& & \, \int d\alpha_0 d\alpha^\prime_0 \rho_s(\pmb{\alpha}_0; 0)\int_{\pmb{\alpha}_0}^{\pmb{\alpha}_t}
\mathcal{D}[\pmb{\alpha}_\tau] e^{iS[\alpha_\tau]-iS[\alpha^\prime_\tau]} \nonumber \\
& & \,\,\, \times 
e^{-\frac{i}{\sqrt 2} \int^t_0 d\tau\left(\mu^\prime_\tau\alpha_\tau + i \nu^\prime_\tau \alpha'_\tau\right)},
\nonumber 
\end{eqnarray}
where $\pmb{\alpha}=(\alpha, \alpha')$ and $\rho(\pmb{\alpha}) \equiv \bra{\alpha'} \rho \ket{\alpha}$.

On the other hand, the trace of $\tilde\rho_B(t)$, governed by Eq.~(\ref{eq:ito_unnorm2}),
can be expressed as
\begin{eqnarray}\label{eq:appc-bnorm}
\text{Tr}\left\{\tilde\rho_B(t)\right\} = \exp\left(-\frac{1}{\sqrt 2} \int^t_0 ds \left( 
\mu(s) +i \nu(t)\right) \mathcal B(t)   \right), \nonumber \\
\end{eqnarray}
where $\mathcal B (t)$ is given by Eq.~(\ref{eq:gaussbfield}).
The exact reduced density matrix is then obtained after formally averaging out the noises
in the following equation,
\begin{eqnarray}
& & \rho_s(\pmb{\alpha}_t; t)  \\
& & \,\, =  \overline{\tilde\rho_s(\pmb{\alpha}_t; t) \text{Tr}\left\{\rho_B(t)\right\}} \nonumber \\
& & \,\, = \int d\alpha_0 d\alpha^\prime_0 \rho_s(\pmb{\alpha}_0; 0)
\int_{\pmb{\alpha}_0}^{\pmb{\alpha}_t} \mathcal{D}[\pmb{\alpha}_\tau]e^{iS[\alpha_\tau]-iS[\alpha^\prime_\tau]} \nonumber \\
& & \,\,\, \times \overline{ 
\exp\left(-\frac{i}{\sqrt 2}\int^t_0 ds  \left\{ \mu^\prime_s\alpha_s + i \nu^\prime_s \alpha'_s
+i \left( \mu_s +i \nu_s\right) \mathcal B_s\right\}  \right)  }, \nonumber
\end{eqnarray}
where an explicit evaluation of the noise average on the last line should yield the standard bosonic bath
influence functional.  To get the influence function, it is useful to contemplate the discretized integrals
for the noise averaging,
\begin{eqnarray}
& & F[\pmb{\alpha}_\tau] = \int \prod_{i} \left[ d\mu_i d\mu^*_i d\nu_i d\nu^*_i 
\left(\frac{\Delta t}{2\pi}\right)^2 e^{-\frac{\Delta t}{2} \left(\vert \mu_i\vert^2 + \vert \nu_i\vert^2 \right) } \right] \nonumber \\
& &\,\, \times \exp\left( -\frac{i}{\sqrt 2} \sum_i \left\{ \mu^*_i \alpha_i 
+ i \nu^*_i \alpha^\prime_i \right\} \right) \nonumber \\
& & \,\, \times\exp\left(\frac{1}{\sqrt 2} \sum_{i\geq j}\left\{ (\mu_i + i\nu_i) (C_{i-j}\mu_j - i C^*_{i-j}\nu_j) \right\}  \right), \nonumber 
\end{eqnarray}
where the bath correlation function $C_{i-j} = C(t_i-t_j)$ is given by
\begin{eqnarray}
C(t) = \sum_{k} \vert g_k \vert^2 \left( 
\cos(\omega_{k}t) \coth\left(\frac{\beta\omega_{k}}{2}\right) - i \sin(\omega_k t) \right). \nonumber \\
\end{eqnarray}
By using the complex-valued Gaussian integral identity,
\begin{eqnarray}
\int dz dz^* e^{-w z z^* + a z + b z^*} = \frac{\pi}{w} e^{-\frac{ab}{w}},
\end{eqnarray}
the standard Feymann-Vermon influence functional is recovered.
The present result is easily generalized when dealing with non-Gaussian baths and $\mathcal{B}(t)$
is potentially characterized by an infinite number of multi-time correlation functions.  
The noise averaging in this general case will give the cumulant expansion of an
influence functional for any bath.

\section{Numerical Illustration of Spin Bath Models With Dual-Fermion based HEOM}\label{app:num}
We present a few numerical results to illustrate the dual-fermion GHE method introduced in this work.
Numerical examples with generalized HEOM approach will be further studied in a separate work, the paper II\cite{hsieh_cao_jcp16}.
We will consider various cases of a pure dephasing model,
\begin{eqnarray}
H= \frac{\omega_0}{2}\sigma^z_0 + \sum_{k>0} \frac{\omega_k}{2}\sigma^z_k
+\sigma^z_0 \sum_{k>0} g_k \sigma^x_k.
\end{eqnarray}
An analytical expression for the off-diagonal matrix element of the reduced
density matrix reads,
\begin{eqnarray}
\langle \uparrow \vert \rho_s(t) \vert \downarrow \rangle = 
\langle \uparrow \vert \rho_s(0) \vert \downarrow \rangle e^{-i\omega_0 t+ \Gamma t},
\end{eqnarray}
with 
\begin{eqnarray}
\Gamma(t) = \sum_{k>0} \ln \left[ 1- \frac{4g_k^2}{\Omega_k^2}\left(1-\cos\Omega_k t\right)\right],
\end{eqnarray}
where $\Omega_k = \omega_k \sqrt{1 + (2g_k/\omega_k)^2}$.

\begin{figure}
\includegraphics[width=\textwidth]{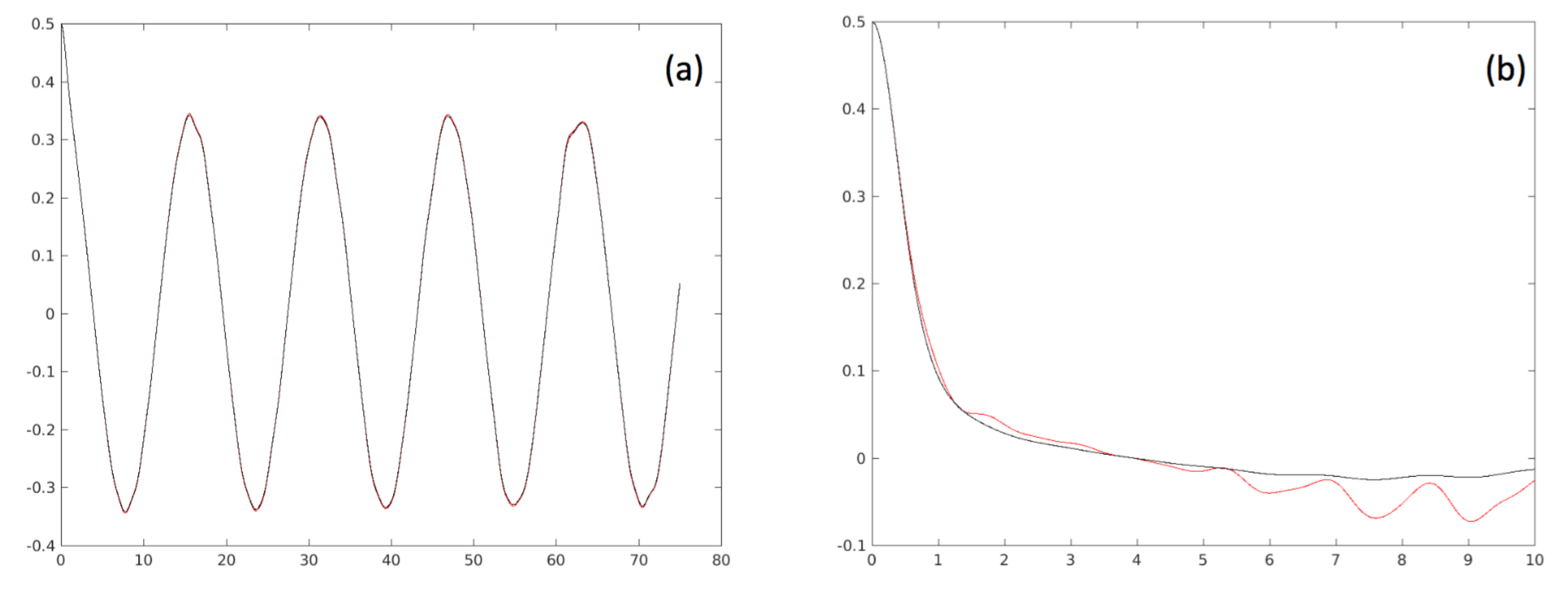}             
\caption{The real part of the off-diagonal matrix element for the RDM.
50 bath spin discretized from an Ohmic spectral density with $\omega_c=1$
$\omega_0=0.4$, and $\alpha$ (Kondo parameter) assumes the value 0.1 (a) and 0.8 (b).
2 hierarchical tiers are used in both cases. Red curves are the numerical results and 
black curves are the exact results.}\label{fig:ohmic}
\end{figure}

First, we consider a 50-spin bath with the parameters $(\omega_k, g_k)$ sampled from the
discretization of an Ohmic bath.  
We use the general dual-fermion GHE scheme, Eq.~(\ref{eq:dualfheom}), to simulate
the off-diagonal matrix element for the density matrix.  Figure \ref{fig:ohmic} shows
the results in the weak coupling (panel a) and the strong coupling (panel b) cases.
Due to each spin is modelled as a bath, it becomes prohibitive to delve into further tiers.
Nevertheless, with a shallow 2-tier hierarchy, the results seem to do
reasonably well in the short-time limit.  

\begin{figure}
\includegraphics[width=\textwidth]{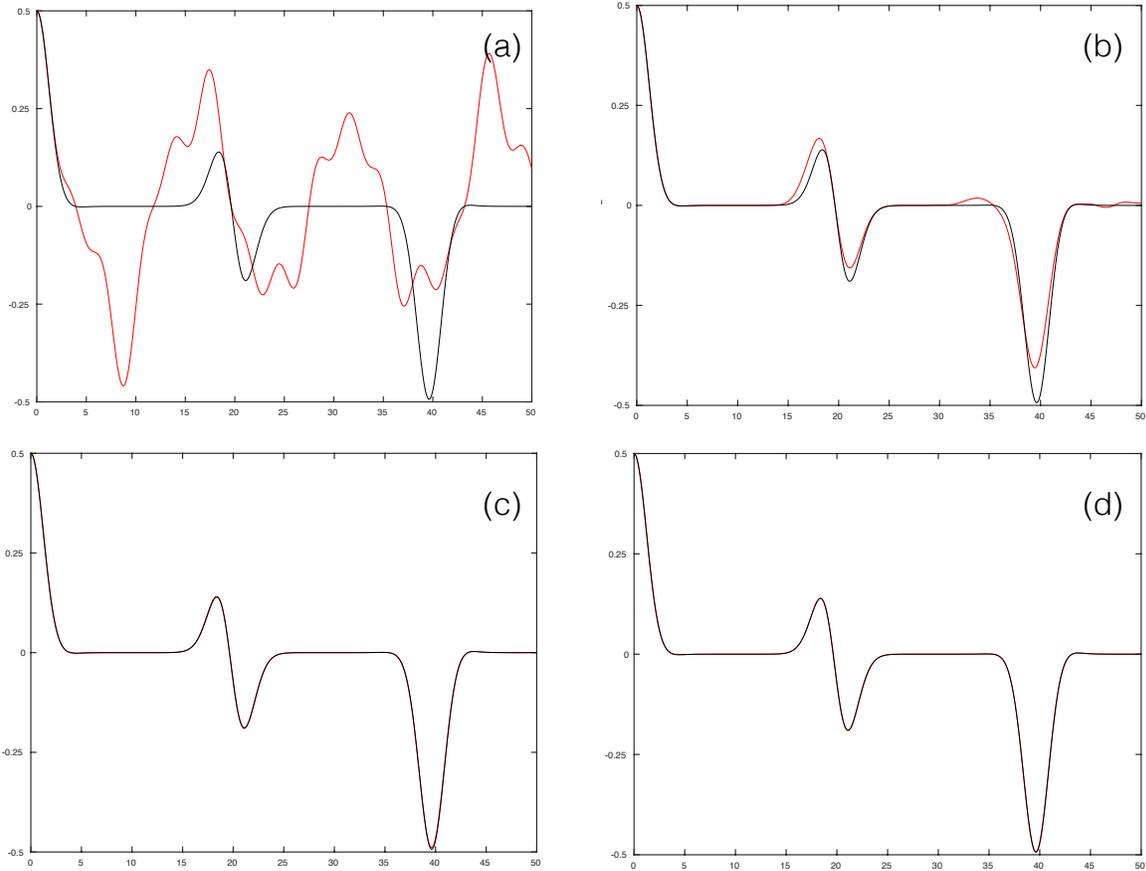}
\caption{The real part of the off-diagonal matrix element for the RDM.
50 bath spins are arranged in a spin-star configuration
with $g_k=0.05$, $\omega_k=0.3$ and $\omega_0=0.4$.  The number of hierarchical tiers (red curves)
are 2 (a), 20 (b), 25(c) and 50 (d), respectively.  The black curves are the exact
results}\label{fig:nuclear}
\end{figure}

In the second case, we consider the spin star model\cite{Ekert12,*wang_guo_eurphys13,*breuer_burgarth_prb04,*petruccione_prb06} where
all the bath spins look identical, i.e. $\omega_k = \omega'$ and $g_k = g'$. 
This is an often used model to analyze spin bath models. 
As shown by the results in Fig.~\ref{fig:nuclear}, it is critical to go deep down
the hierarchical tiers in order to recover the correct quantum dissipations.  One can only 
generate this many tiers through compressing the auxiliary density matrices as 
in Eq.~(\ref{eq:dualfc}).  This second example illustrates the kind of scenarios where dual-fermion
GHE could provide an accurate account of quantum dynamics induced by a spin bath.

\begin{thebibliography}{45}
\expandafter\ifx\csname natexlab\endcsname\relax\def\natexlab#1{#1}\fi
\expandafter\ifx\csname bibnamefont\endcsname\relax
  \def\bibnamefont#1{#1}\fi
\expandafter\ifx\csname bibfnamefont\endcsname\relax
  \def\bibfnamefont#1{#1}\fi
\expandafter\ifx\csname citenamefont\endcsname\relax
  \def\citenamefont#1{#1}\fi
\expandafter\ifx\csname url\endcsname\relax
  \def\url#1{\texttt{#1}}\fi
\expandafter\ifx\csname urlprefix\endcsname\relax\def\urlprefix{URL }\fi
\providecommand{\bibinfo}[2]{#2}
\providecommand{\eprint}[2][]{\url{#2}}

\bibitem[{\citenamefont{Leggett et~al.}(1987)\citenamefont{Leggett,
  Chakravarty, Dorsey, Fisher, Garg, and Zwerger}}]{Leggett:1987wk}
\bibinfo{author}{\bibfnamefont{A.~J.} \bibnamefont{Leggett}},
  \bibinfo{author}{\bibfnamefont{S.}~\bibnamefont{Chakravarty}},
  \bibinfo{author}{\bibfnamefont{A.~T.} \bibnamefont{Dorsey}},
  \bibinfo{author}{\bibfnamefont{M.~P.~A.} \bibnamefont{Fisher}},
  \bibinfo{author}{\bibfnamefont{A.}~\bibnamefont{Garg}}, \bibnamefont{and}
  \bibinfo{author}{\bibfnamefont{W.}~\bibnamefont{Zwerger}},
  \bibinfo{journal}{Rev. Mod. Phys.} \textbf{\bibinfo{volume}{59}},
  \bibinfo{pages}{1} (\bibinfo{year}{1987}).

\bibitem[{\citenamefont{Breuer and Petruccione}(2002)}]{breuer:book}
\bibinfo{author}{\bibfnamefont{H.-P.} \bibnamefont{Breuer}} \bibnamefont{and}
  \bibinfo{author}{\bibfnamefont{F.}~\bibnamefont{Petruccione}},
  \emph{\bibinfo{title}{The Theory of Open Quantum Systems}}
  (\bibinfo{publisher}{Oxford University Press}, \bibinfo{address}{New York},
  \bibinfo{year}{2002}).

\bibitem[{\citenamefont{Weiss}(2012)}]{weiss:book}
\bibinfo{author}{\bibfnamefont{U.}~\bibnamefont{Weiss}},
  \emph{\bibinfo{title}{Quantum Dissipative Systems, 3rd ed., Series in Modern
  Condensed Matter Physics Vol. 13}} (\bibinfo{publisher}{World Scientific},
  \bibinfo{address}{Singapore}, \bibinfo{year}{2012}).

\bibitem[{\citenamefont{Wang and Thoss}(2007)}]{Wang:2007eu}
\bibinfo{author}{\bibfnamefont{H.}~\bibnamefont{Wang}} \bibnamefont{and}
  \bibinfo{author}{\bibfnamefont{M.}~\bibnamefont{Thoss}}, \bibinfo{journal}{J.
  Phys. Chem. A} \textbf{\bibinfo{volume}{111}}, \bibinfo{pages}{10369}
  (\bibinfo{year}{2007}).

\bibitem[{\citenamefont{L{\'o}pez-L{\'o}pez
  et~al.}(2011)\citenamefont{L{\'o}pez-L{\'o}pez, Martinazzo, and
  Nest}}]{LopezLopez:2011en}
\bibinfo{author}{\bibfnamefont{S.}~\bibnamefont{L{\'o}pez-L{\'o}pez}},
  \bibinfo{author}{\bibfnamefont{R.}~\bibnamefont{Martinazzo}},
  \bibnamefont{and} \bibinfo{author}{\bibfnamefont{M.}~\bibnamefont{Nest}},
  \bibinfo{journal}{J. Chem. Phys.} \textbf{\bibinfo{volume}{134}},
  \bibinfo{pages}{094102} (\bibinfo{year}{2011}).

\bibitem[{\citenamefont{Kryvohuz and Cao}(2005)}]{Kryvohuz:2005jf}
\bibinfo{author}{\bibfnamefont{M.}~\bibnamefont{Kryvohuz}} \bibnamefont{and}
  \bibinfo{author}{\bibfnamefont{J.}~\bibnamefont{Cao}},
  \bibinfo{journal}{Phys. Rev. Lett.} \textbf{\bibinfo{volume}{95}},
  \bibinfo{pages}{180405} (\bibinfo{year}{2005}).

\bibitem[{\citenamefont{Wu and Cao}(2001)}]{wu_cao_jcp01}
\bibinfo{author}{\bibfnamefont{J.}~\bibnamefont{Wu}} \bibnamefont{and}
  \bibinfo{author}{\bibfnamefont{J.}~\bibnamefont{Cao}}, \bibinfo{journal}{J.
  Chem. Phys.} \textbf{\bibinfo{volume}{115}}, \bibinfo{pages}{5381}
  (\bibinfo{year}{2001}).

\bibitem[{\citenamefont{Cao and Voth}(1995)}]{cao_voth_jcp95}
\bibinfo{author}{\bibfnamefont{J.}~\bibnamefont{Cao}} \bibnamefont{and}
  \bibinfo{author}{\bibfnamefont{G.~A.} \bibnamefont{Voth}},
  \bibinfo{journal}{J. Chem. Phys.} \textbf{\bibinfo{volume}{102}},
  \bibinfo{pages}{3337} (\bibinfo{year}{1995}).

\bibitem[{\citenamefont{Makri}(1999)}]{makri99}
\bibinfo{author}{\bibfnamefont{N.}~\bibnamefont{Makri}}, \bibinfo{journal}{J.
  Phys. Chem. B} \textbf{\bibinfo{volume}{103}}, \bibinfo{pages}{2823}
  (\bibinfo{year}{1999}).

\bibitem[{\citenamefont{Kotliar et~al.}(2006)\citenamefont{Kotliar, Savrasov,
  Haule, Oudovenko, Parcollet, and Marianetti}}]{dmft06}
\bibinfo{author}{\bibfnamefont{G.}~\bibnamefont{Kotliar}},
  \bibinfo{author}{\bibfnamefont{S.~Y.} \bibnamefont{Savrasov}},
  \bibinfo{author}{\bibfnamefont{K.}~\bibnamefont{Haule}},
  \bibinfo{author}{\bibfnamefont{V.~S.} \bibnamefont{Oudovenko}},
  \bibinfo{author}{\bibfnamefont{O.}~\bibnamefont{Parcollet}},
  \bibnamefont{and} \bibinfo{author}{\bibfnamefont{C.~A.}
  \bibnamefont{Marianetti}}, \bibinfo{journal}{Rev. Mod. Phys.}
  \textbf{\bibinfo{volume}{78}}, \bibinfo{pages}{865} (\bibinfo{year}{2006}).

\bibitem[{\citenamefont{Knizia and Chan}(2012)}]{dmet12}
\bibinfo{author}{\bibfnamefont{G.}~\bibnamefont{Knizia}} \bibnamefont{and}
  \bibinfo{author}{\bibfnamefont{G.~K.-L.} \bibnamefont{Chan}},
  \bibinfo{journal}{Phys. Rev. Lett.} \textbf{\bibinfo{volume}{109}},
  \bibinfo{pages}{186404} (\bibinfo{year}{2012}).

\bibitem[{\citenamefont{Hsieh et~al.}(2012)\citenamefont{Hsieh, Shim,
  Korkusinski, and Hawrylak}}]{hsieh_2012}
\bibinfo{author}{\bibfnamefont{C.-Y.} \bibnamefont{Hsieh}},
  \bibinfo{author}{\bibfnamefont{Y.-P.} \bibnamefont{Shim}},
  \bibinfo{author}{\bibfnamefont{M.}~\bibnamefont{Korkusinski}},
  \bibnamefont{and} \bibinfo{author}{\bibfnamefont{P.}~\bibnamefont{Hawrylak}},
  \bibinfo{journal}{Rep. Prog. Phys.} \textbf{\bibinfo{volume}{75}},
  \bibinfo{pages}{114501} (\bibinfo{year}{2012}).

\bibitem[{\citenamefont{Stockburger and Grabert}(2002)}]{Stockburger:2002em}
\bibinfo{author}{\bibfnamefont{J.}~\bibnamefont{Stockburger}} \bibnamefont{and}
  \bibinfo{author}{\bibfnamefont{H.}~\bibnamefont{Grabert}},
  \bibinfo{journal}{Phys. Rev. Lett.} \textbf{\bibinfo{volume}{88}},
  \bibinfo{pages}{170407} (\bibinfo{year}{2002}).

\bibitem[{\citenamefont{Lacroix}(2005)}]{Lacroix:2005in}
\bibinfo{author}{\bibfnamefont{D.}~\bibnamefont{Lacroix}},
  \bibinfo{journal}{Phys. Rev. A} \textbf{\bibinfo{volume}{72}},
  \bibinfo{pages}{013805} (\bibinfo{year}{2005}).

\bibitem[{\citenamefont{Shao}(2004)}]{Shao:2004et}
\bibinfo{author}{\bibfnamefont{J.}~\bibnamefont{Shao}}, \bibinfo{journal}{J.
  Chem. Phys.} \textbf{\bibinfo{volume}{120}}, \bibinfo{pages}{5053}
  (\bibinfo{year}{2004}).

\bibitem[{\citenamefont{Stockburger}(2016)}]{stockburger_epl2016}
\bibinfo{author}{\bibfnamefont{J.}~\bibnamefont{Stockburger}},
  \bibinfo{journal}{Eur. Phys. Lett.} \textbf{\bibinfo{volume}{115}},
  \bibinfo{pages}{40010} (\bibinfo{year}{2016}).

\bibitem[{\citenamefont{Wiedmann et~al.}(2016)\citenamefont{Wiedmann,
  Stockburger, and Ankerhold}}]{wiedmann_stockburger_pra2016}
\bibinfo{author}{\bibfnamefont{M.}~\bibnamefont{Wiedmann}},
  \bibinfo{author}{\bibfnamefont{J.~T.} \bibnamefont{Stockburger}},
  \bibnamefont{and}
  \bibinfo{author}{\bibfnamefont{J.}~\bibnamefont{Ankerhold}},
  \bibinfo{journal}{Phys. Rev. A} \textbf{\bibinfo{volume}{94}},
  \bibinfo{pages}{052137} (\bibinfo{year}{2016}).

\bibitem[{\citenamefont{Zhou et~al.}(2016)\citenamefont{Zhou, Chen, Yu, and
  You}}]{zhou_chen_pra2016}
\bibinfo{author}{\bibfnamefont{Z.-Y.} \bibnamefont{Zhou}},
  \bibinfo{author}{\bibfnamefont{M.}~\bibnamefont{Chen}},
  \bibinfo{author}{\bibfnamefont{T.}~\bibnamefont{Yu}}, \bibnamefont{and}
  \bibinfo{author}{\bibfnamefont{J.~Q.} \bibnamefont{You}},
  \bibinfo{journal}{Phys. Rev. A} \textbf{\bibinfo{volume}{93}},
  \bibinfo{pages}{022105} (\bibinfo{year}{2016}).

\bibitem[{\citenamefont{Moix et~al.}(2012)\citenamefont{Moix, Zhao, and
  Cao}}]{moix_prb2012}
\bibinfo{author}{\bibfnamefont{J.~M.} \bibnamefont{Moix}},
  \bibinfo{author}{\bibfnamefont{Y.}~\bibnamefont{Zhao}}, \bibnamefont{and}
  \bibinfo{author}{\bibfnamefont{J.}~\bibnamefont{Cao}},
  \bibinfo{journal}{Phys. Rev. B} \textbf{\bibinfo{volume}{85}},
  \bibinfo{pages}{115412} (\bibinfo{year}{2012}).

\bibitem[{\citenamefont{Moix et~al.}(2015)\citenamefont{Moix, Ma, and
  Cao}}]{Moix:2015ei}
\bibinfo{author}{\bibfnamefont{J.~M.} \bibnamefont{Moix}},
  \bibinfo{author}{\bibfnamefont{J.}~\bibnamefont{Ma}}, \bibnamefont{and}
  \bibinfo{author}{\bibfnamefont{J.}~\bibnamefont{Cao}}, \bibinfo{journal}{J.
  Chem. Phys.} \textbf{\bibinfo{volume}{142}}, \bibinfo{pages}{094108}
  (\bibinfo{year}{2015}).

\bibitem[{\citenamefont{Moix and Cao}(2013)}]{Moix:2013jb}
\bibinfo{author}{\bibfnamefont{J.~M.} \bibnamefont{Moix}} \bibnamefont{and}
  \bibinfo{author}{\bibfnamefont{J.}~\bibnamefont{Cao}}, \bibinfo{journal}{J.
  Chem. Phys.} \textbf{\bibinfo{volume}{139}}, \bibinfo{pages}{134106}
  (\bibinfo{year}{2013}).

\bibitem[{\citenamefont{Chen and You}(2013)}]{Chen_You_pra2013}
\bibinfo{author}{\bibfnamefont{M.}~\bibnamefont{Chen}} \bibnamefont{and}
  \bibinfo{author}{\bibfnamefont{J.~Q.} \bibnamefont{You}},
  \bibinfo{journal}{Phys. Rev. A} \textbf{\bibinfo{volume}{87}},
  \bibinfo{pages}{052108} (\bibinfo{year}{2013}).

\bibitem[{\citenamefont{Li et~al.}(2011)\citenamefont{Li, Shao, and
  Wang}}]{Li:2011dj}
\bibinfo{author}{\bibfnamefont{H.}~\bibnamefont{Li}},
  \bibinfo{author}{\bibfnamefont{J.}~\bibnamefont{Shao}}, \bibnamefont{and}
  \bibinfo{author}{\bibfnamefont{S.}~\bibnamefont{Wang}},
  \bibinfo{journal}{Phys. Rev. E} \textbf{\bibinfo{volume}{84}},
  \bibinfo{pages}{051112} (\bibinfo{year}{2011}).

\bibitem[{\citenamefont{de~Vega}(2015)}]{vega2015}
\bibinfo{author}{\bibfnamefont{I.}~\bibnamefont{de~Vega}},
  \bibinfo{journal}{J.Phys. A} \textbf{\bibinfo{volume}{48}},
  \bibinfo{pages}{145202} (\bibinfo{year}{2015}).

\bibitem[{\citenamefont{Zhou et~al.}(2007)\citenamefont{Zhou, Yan, and
  Shao}}]{Zhou:2007fx}
\bibinfo{author}{\bibfnamefont{Y.}~\bibnamefont{Zhou}},
  \bibinfo{author}{\bibfnamefont{Y.}~\bibnamefont{Yan}}, \bibnamefont{and}
  \bibinfo{author}{\bibfnamefont{J.}~\bibnamefont{Shao}},
  \bibinfo{journal}{Europhys. Lett.} \textbf{\bibinfo{volume}{72}},
  \bibinfo{pages}{334} (\bibinfo{year}{2007}).

\bibitem[{\citenamefont{Yan}(2014)}]{Yan:2014cc}
\bibinfo{author}{\bibfnamefont{Y.}~\bibnamefont{Yan}}, \bibinfo{journal}{J.
  Chem. Phys.} \textbf{\bibinfo{volume}{140}}, \bibinfo{pages}{054105}
  (\bibinfo{year}{2014}).

\bibitem[{\citenamefont{Tanimura}(2006)}]{Tanimura:2006ga}
\bibinfo{author}{\bibfnamefont{Y.}~\bibnamefont{Tanimura}},
  \bibinfo{journal}{J. Phys. Soc. Jpn.} \textbf{\bibinfo{volume}{75}},
  \bibinfo{pages}{082001} (\bibinfo{year}{2006}).

\bibitem[{\citenamefont{Ishizaki and
  Tanimura}(2005)}]{ishizaki_tanimura_jpsp2005}
\bibinfo{author}{\bibfnamefont{A.}~\bibnamefont{Ishizaki}} \bibnamefont{and}
  \bibinfo{author}{\bibfnamefont{Y.}~\bibnamefont{Tanimura}},
  \bibinfo{journal}{J. Phys. Soc. Jpn.} \textbf{\bibinfo{volume}{74}},
  \bibinfo{pages}{3131} (\bibinfo{year}{2005}).

\bibitem[{\citenamefont{Tang et~al.}(2015)\citenamefont{Tang, Ouyang, Gong,
  Wang, and Wu}}]{Tang:2016gh}
\bibinfo{author}{\bibfnamefont{Z.}~\bibnamefont{Tang}},
  \bibinfo{author}{\bibfnamefont{X.}~\bibnamefont{Ouyang}},
  \bibinfo{author}{\bibfnamefont{Z.}~\bibnamefont{Gong}},
  \bibinfo{author}{\bibfnamefont{H.}~\bibnamefont{Wang}}, \bibnamefont{and}
  \bibinfo{author}{\bibfnamefont{J.}~\bibnamefont{Wu}}, \bibinfo{journal}{J.
  Chem. Phys.} \textbf{\bibinfo{volume}{143}}, \bibinfo{pages}{224112}
  (\bibinfo{year}{2015}).

\bibitem[{\citenamefont{Ye et~al.}(2016)\citenamefont{Ye, Wang, Hou, Xu, Zheng,
  and Yan}}]{yyj_wiley16}
\bibinfo{author}{\bibfnamefont{L.}~\bibnamefont{Ye}},
  \bibinfo{author}{\bibfnamefont{X.}~\bibnamefont{Wang}},
  \bibinfo{author}{\bibfnamefont{D.}~\bibnamefont{Hou}},
  \bibinfo{author}{\bibfnamefont{R.-X.} \bibnamefont{Xu}},
  \bibinfo{author}{\bibfnamefont{X.}~\bibnamefont{Zheng}}, \bibnamefont{and}
  \bibinfo{author}{\bibfnamefont{Y.}~\bibnamefont{Yan}},
  \bibinfo{journal}{WIREs Comput Mol Sci} \textbf{\bibinfo{volume}{6}},
  \bibinfo{pages}{608} (\bibinfo{year}{2016}).

\bibitem[{\citenamefont{Prokofiev and {Stamp, P. C.
  E.}}(2000)}]{Prokofiev:420342}
\bibinfo{author}{\bibfnamefont{N.~V.} \bibnamefont{Prokofiev}}
  \bibnamefont{and} \bibinfo{author}{\bibnamefont{{Stamp, P. C. E.}}},
  \bibinfo{journal}{Rep. Prog. Phys.} \textbf{\bibinfo{volume}{63}},
  \bibinfo{pages}{669} (\bibinfo{year}{2000}).

\bibitem[{\citenamefont{Kloeffel and Loss}(2013)}]{Kloeffel:2013eg}
\bibinfo{author}{\bibfnamefont{C.}~\bibnamefont{Kloeffel}} \bibnamefont{and}
  \bibinfo{author}{\bibfnamefont{D.}~\bibnamefont{Loss}},
  \bibinfo{journal}{Anuu. Rev. Conden. Ma. P.} \textbf{\bibinfo{volume}{4}},
  \bibinfo{pages}{51} (\bibinfo{year}{2013}).

\bibitem[{\citenamefont{Hsieh and Cao}(2016)}]{hsieh_cao_jcp16}
\bibinfo{author}{\bibfnamefont{C.-Y.} \bibnamefont{Hsieh}} \bibnamefont{and}
  \bibinfo{author}{\bibfnamefont{J.S.}~\bibnamefont{Cao}}  \bibinfo{journal}{Paper II}
  (\bibinfo{year}{2016}).

\bibitem[{\citenamefont{Cao et~al.}(1996)\citenamefont{Cao, Ungar, and
  Voth}}]{cao_ungar_jcp96}
\bibinfo{author}{\bibfnamefont{J.}~\bibnamefont{Cao}},
  \bibinfo{author}{\bibfnamefont{L.~W.} \bibnamefont{Ungar}}, \bibnamefont{and}
  \bibinfo{author}{\bibfnamefont{G.~A.} \bibnamefont{Voth}},
  \bibinfo{journal}{J. Chem. Phys.} \textbf{\bibinfo{volume}{104}},
  \bibinfo{pages}{4189} (\bibinfo{year}{1996}).

\bibitem[{\citenamefont{Di{\'o}si et~al.}(1998)\citenamefont{Di{\'o}si, Gisin,
  and Strunz}}]{Diosi:1998px}
\bibinfo{author}{\bibfnamefont{L.}~\bibnamefont{Di{\'o}si}},
  \bibinfo{author}{\bibfnamefont{N.}~\bibnamefont{Gisin}}, \bibnamefont{and}
  \bibinfo{author}{\bibfnamefont{W.~T.} \bibnamefont{Strunz}},
  \bibinfo{journal}{Phys. Rev. A} \textbf{\bibinfo{volume}{58}},
  \bibinfo{pages}{1699} (\bibinfo{year}{1998}).

\bibitem[{\citenamefont{Zhong et~al.}(2014)\citenamefont{Zhong, Zhao, and
  Cao}}]{zhong_cao_14}
\bibinfo{author}{\bibfnamefont{X.}~\bibnamefont{Zhong}},
  \bibinfo{author}{\bibfnamefont{Y.}~\bibnamefont{Zhao}}, \bibnamefont{and}
  \bibinfo{author}{\bibfnamefont{J.}~\bibnamefont{Cao}}, \bibinfo{journal}{New
  J. Phys.} \textbf{\bibinfo{volume}{16}}, \bibinfo{pages}{045009}
  (\bibinfo{year}{2014}).

\bibitem[{\citenamefont{Zhao et~al.}(2012)\citenamefont{Zhao, Shi, Wu, and
  Yu}}]{Zhao_Yu_pra2012}
\bibinfo{author}{\bibfnamefont{X.}~\bibnamefont{Zhao}},
  \bibinfo{author}{\bibfnamefont{W.}~\bibnamefont{Shi}},
  \bibinfo{author}{\bibfnamefont{L.-A.} \bibnamefont{Wu}}, \bibnamefont{and}
  \bibinfo{author}{\bibfnamefont{T.}~\bibnamefont{Yu}}, \bibinfo{journal}{Phys.
  Rev. A} \textbf{\bibinfo{volume}{86}}, \bibinfo{pages}{032116}
  (\bibinfo{year}{2012}).

\bibitem[{\citenamefont{Suess et~al.}(2015)\citenamefont{Suess, Strunz, and
  Eisfeld}}]{Suess:2015iu}
\bibinfo{author}{\bibfnamefont{D.}~\bibnamefont{Suess}},
  \bibinfo{author}{\bibfnamefont{W.~T.} \bibnamefont{Strunz}},
  \bibnamefont{and} \bibinfo{author}{\bibfnamefont{A.}~\bibnamefont{Eisfeld}},
  \bibinfo{journal}{J. Stat. Phys.} \textbf{\bibinfo{volume}{159}},
  \bibinfo{pages}{1408} (\bibinfo{year}{2015}).

\bibitem[{\citenamefont{Suess et~al.}(2014)\citenamefont{Suess, Eisfeld, and
  Strunz}}]{Suess:2014gz}
\bibinfo{author}{\bibfnamefont{D.}~\bibnamefont{Suess}},
  \bibinfo{author}{\bibfnamefont{A.}~\bibnamefont{Eisfeld}}, \bibnamefont{and}
  \bibinfo{author}{\bibfnamefont{W.~T.} \bibnamefont{Strunz}},
  \bibinfo{journal}{Phys. Rev. Lett.} \textbf{\bibinfo{volume}{113}},
  \bibinfo{pages}{150403} (\bibinfo{year}{2014}).

\bibitem[{\citenamefont{Song et~al.}(2016)\citenamefont{Song, Song, and
  Shi}}]{qiang2016}
\bibinfo{author}{\bibfnamefont{K.}~\bibnamefont{Song}},
  \bibinfo{author}{\bibfnamefont{L.}~\bibnamefont{Song}}, \bibnamefont{and}
  \bibinfo{author}{\bibfnamefont{Q.}~\bibnamefont{Shi}}, \bibinfo{journal}{J.
  Chem. Phys.} \textbf{\bibinfo{volume}{144}}, \bibinfo{eid}{224105}
  (\bibinfo{year}{2016}).

\bibitem[{\citenamefont{Young and Grace}(2013)}]{young_prl2013}
\bibinfo{author}{\bibfnamefont{K.~C.} \bibnamefont{Young}} \bibnamefont{and}
  \bibinfo{author}{\bibfnamefont{M.~D.} \bibnamefont{Grace}},
  \bibinfo{journal}{Phys. Rev. Lett.} \textbf{\bibinfo{volume}{110}},
  \bibinfo{pages}{110402} (\bibinfo{year}{2013}).

\bibitem[{\citenamefont{Sinayskiy et~al.}(2012)\citenamefont{Sinayskiy, Marais,
  Petruccione, and Ekert}}]{Ekert12}
\bibinfo{author}{\bibfnamefont{I.}~\bibnamefont{Sinayskiy}},
  \bibinfo{author}{\bibfnamefont{A.}~\bibnamefont{Marais}},
  \bibinfo{author}{\bibfnamefont{F.}~\bibnamefont{Petruccione}},
  \bibnamefont{and} \bibinfo{author}{\bibfnamefont{A.}~\bibnamefont{Ekert}},
  \bibinfo{journal}{Phys. Rev. Lett.} \textbf{\bibinfo{volume}{108}},
  \bibinfo{pages}{020602} (\bibinfo{year}{2012}).

\bibitem[{\citenamefont{Wang et~al.}(2013)\citenamefont{Wang, Guo, and
  Zhou}}]{wang_guo_eurphys13}
\bibinfo{author}{\bibfnamefont{Z.}~\bibnamefont{Wang}},
  \bibinfo{author}{\bibfnamefont{Y.}~\bibnamefont{Guo}}, \bibnamefont{and}
  \bibinfo{author}{\bibfnamefont{D.~L.} \bibnamefont{Zhou}},
  \bibinfo{journal}{Eur. Phys. J. D} \textbf{\bibinfo{volume}{67}},
  \bibinfo{pages}{218} (\bibinfo{year}{2013}).

\bibitem[{\citenamefont{Breuer et~al.}(2004)\citenamefont{Breuer, Burgarth, and
  Petruccione}}]{breuer_burgarth_prb04}
\bibinfo{author}{\bibfnamefont{H.-P.} \bibnamefont{Breuer}},
  \bibinfo{author}{\bibfnamefont{D.}~\bibnamefont{Burgarth}}, \bibnamefont{and}
  \bibinfo{author}{\bibfnamefont{F.}~\bibnamefont{Petruccione}},
  \bibinfo{journal}{Phys. Rev. B} p. \bibinfo{pages}{045323}
  (\bibinfo{year}{2004}).

\bibitem[{\citenamefont{Hamdouni et~al.}(2006)\citenamefont{Hamdouni, Fannes,
  and Petruccione}}]{petruccione_prb06}
\bibinfo{author}{\bibfnamefont{Y.}~\bibnamefont{Hamdouni}},
  \bibinfo{author}{\bibfnamefont{M.}~\bibnamefont{Fannes}}, \bibnamefont{and}
  \bibinfo{author}{\bibfnamefont{F.}~\bibnamefont{Petruccione}},
  \bibinfo{journal}{Phys. Rev. B} \textbf{\bibinfo{volume}{73}},
  \bibinfo{pages}{245323} (\bibinfo{year}{2006}).

\end{thebibliography}

\end{document}